\documentclass[a4paper,openany, 12pt]{book}
\usepackage[text={15.5cm,23cm},centering]{geometry}

\newcommand\savemathcal[1]{%
  \expandafter\newsavebox\csname mc#1content\endcsname%
  \expandafter\savebox\csname mc#1content\endcsname{$\mathcal{#1}$}%
  \expandafter\newcommand\csname mc#1\endcsname{%
    \expandafter\usebox\expandafter{\csname mc#1content\endcsname}}%
}
\newcommand\altmathcal[1]{\csname mc#1\endcsname}
\savemathcal{Z}
\savemathcal{B}

\usepackage{mathptmx}
\usepackage{graphicx}
\usepackage{chapterbib} %remember to bibtex each chapter before final compilation
\usepackage{subfigure}
\usepackage{color}
\usepackage{amsmath}
\usepackage{amssymb}
\usepackage[colorlinks=true, linkcolor=blue]{hyperref}
\usepackage[english]{babel}  %For Vibor's institution characters

\title{The cosmic 21-cm revolution: charting the first billion years of our Universe}
\author{Andrei Mesinger}

\begin{document}
\frontmatter
%\maketitle
%\tableofcontents

%\include{Mesinger/Preface}
%\include{Mesinger/Author}
%\include{Contributors} %only needed for edited books

\mainmatter

\newcommand*{\dt}[1]{%
  \accentset{\mbox{\large\bfseries .}}{#1}}
\newcommand*{\ddt}[1]{%
  \accentset{\mbox{\large\bfseries .\hspace{-0.25ex}.}}{#1}}

% Densities
\newcommand{\nH}{n_{\text{H}}}
\newcommand{\nHe}{n_{\text{He}}}
\newcommand{\nHbar}{\bar{n}_{\text{H}}^0}
\newcommand{\nHebar}{\bar{n}_{\text{He}}^0}
\newcommand{\nbbar}{\bar{n}_{\text{b}}^0}
\newcommand{\rhobbar}{\bar{rho}_{\text{b}}^0}

% Hydrogen and helium ions - don't add $$
\newcommand{\HI}{\text{H} {\textsc{i}}}
\newcommand{\HII}{\text{H} {\textsc{ii}}}
\newcommand{\HeI}{\text{He} {\textsc{i}}}
\newcommand{\HeII}{\text{He} {\textsc{ii}}}
\newcommand{\HeIII}{\text{He} {\textsc{iii}}}
\newcommand{\Htwo}{\text{H}_2}
\newcommand{\Hminus}{\text{H}^{-}}
\newcommand{\Hatom}{\text{H}}
\newcommand{\xibar}{\overline{x}_i}
\newcommand{\QHII}{Q_{\HII}}

\newcommand{\xipr}{x_i^{\prime}}
\newcommand{\Emin}{E_{\min}}

\newcommand{\Nion}{N_{\text{ion}}}
\newcommand{\Nlw}{N_{\text{LW}}}
\newcommand{\zetaI}{\zeta_{\text{ion}}}
\newcommand{\zetaX}{\zeta_X}
\newcommand{\zetaA}{\zeta_{\alpha}}

\newcommand{\nuLL}{\nu_{\text{LL}}}
\newcommand{\nuLya}{\nu_{\alpha}}

% Number densities of common ions
\newcommand{\nHI}{n_{\text{H } \textsc{i}}}
\newcommand{\nHII}{n_{\text{H } \textsc{ii}}}
\newcommand{\nHeI}{n_{\text{He } \textsc{i}}}
\newcommand{\nHeII}{n_{\text{He } \textsc{ii}}}
\newcommand{\nHeIII}{n_{\text{He } \textsc{iii}}}
\newcommand{\nel}{n_{\text{e}}}  
\newcommand{\ntot}{n_{\text{tot}}}

% Species fractions
\newcommand{\xHI}{x_{\text{H } \textsc{i}}}
\newcommand{\xHII}{x_{\text{H } \textsc{ii}}}
\newcommand{\xHeI}{x_{\text{He } \textsc{i}}}
\newcommand{\xHeII}{x_{\text{He } \textsc{ii}}}
\newcommand{\xHeIII}{x_{\text{He } \textsc{iii}}}

% Ionization & Recombination coefficients
\newcommand{\ionHI}{\Gamma_{\text{H } \textsc{i}}}
\newcommand{\ionHeI}{\Gamma_{\text{He } \textsc{i}}}
\newcommand{\ionHeII}{\Gamma_{\text{He } \textsc{ii}}}
\newcommand{\ionsecHI}{\gamma_{\text{H } \textsc{i}}}
\newcommand{\ionsecHeI}{\gamma_{\text{He } \textsc{i}}}
\newcommand{\ionsecHeII}{\gamma_{\text{He } \textsc{ii}}}
\newcommand{\ioncollHI}{\beta_{\text{H } \textsc{i}}}
\newcommand{\ioncollHeI}{\beta_{\text{He } \textsc{i}}}
\newcommand{\ioncollHeII}{\beta_{\text{He } \textsc{ii}}}
\newcommand{\recHII}{\alpha_{\text{H } \textsc{ii}}}
\newcommand{\recHeII}{\alpha_{\text{He } \textsc{ii}}}
\newcommand{\recHeIII}{\alpha_{\text{He } \textsc{iii}}}

\newcommand{\xiHeII}{\xi_{\text{He} \textsc{ii}}}

% Heating rate coefficients
\newcommand{\heatHI}{\mathcal{H}_{\text{H } \textsc{i}}}
\newcommand{\heatHeI}{\mathcal{H}_{\text{He } \textsc{i}}}
\newcommand{\heatHeII}{\mathcal{H}_{\text{He } \textsc{ii}}}

% Cooling rate coefficients
\newcommand{\cooldielHeII}{\omega_{\text{He } \textsc{ii}}}

% Phi and Psi
\newcommand{\PhiHI}{\Phi_{\text{H } \textsc{i}}}
\newcommand{\PhiHeI}{\Phi_{\text{He } \textsc{i}}}
\newcommand{\PhiHeII}{\Phi_{\text{He } \textsc{ii}}}
\newcommand{\PsiHI}{\Psi_{\text{H } \textsc{i}}}
\newcommand{\PsiHeI}{\Psi_{\text{He } \textsc{i}}}
\newcommand{\PsiHeII}{\Psi_{\text{He } \textsc{ii}}}

% BH stuff
\newcommand{\fduty}{f_{\text{duty}}}
\newcommand{\Cedd}{C_{\text{edd}}}
\newcommand{\tedd}{t_{\text{edd}}}
\newcommand{\Mbh}{M_{\bullet}}

\newcommand{\SFR}{\dot{M}_{\ast}}
\newcommand{\MAR}{\dot{M}_{b}}
\newcommand{\SFE}{f_{\ast}}
\newcommand{\Tmin}{T_{\min}}

% Random
\newcommand{\zprime}{z^{\prime}}
\newcommand{\dprime}{\prime\prime}
\newcommand{\fstar}{f_{\ast}}
\newcommand{\fstarbh}{\tilde{\fstar}}
\newcommand{\fbh}{f_{\bullet}}
\newcommand{\fcoll}{f_{\text{coll}}}
\newcommand{\dfcolldz}{\frac{df_{\text{coll}}}{dz}}
\newcommand{\dfcolldt}{\frac{df_{\text{coll}}}{dt}}
\newcommand{\dfcolldzbh}{\frac{d\tilde{f}_{\text{coll}}}{dz}}
\newcommand{\dfcolldtbh}{\frac{d\tilde{f}_{\text{coll}}}{dt}}
\newcommand{\mmin}{m_{\text{min}}}
\newcommand{\rhobh}{\rho_{\bullet}}
\newcommand{\rhobhdot}{\dt{\rho}_{\bullet}}
\newcommand{\rhostar}{\rho_{\ast}}
\newcommand{\rhostardot}{\dt{\rho}_{\ast}}
\newcommand{\rhostarbhdot}{\dt{\rho}_{\ast\bullet}}
\newcommand{\rhom}{\rho_m}
\newcommand{\fstardegen}{f_{\ast \bullet}}
\newcommand{\Ndot}{\dot{N}_{\text{ion}}}
\newcommand{\fesc}{f_{\text{esc}}}
\newcommand{\nmax}{n_{\text{max}}}
\newcommand{\frec}{f_{\text{rec}}}
\newcommand{\frecn}{f_{\text{rec}}^{n}}
\newcommand{\frecbar}{\overline{f}_{\text{rec}}}
\newcommand{\Msun}{M_{\odot}}

\newcommand{\SFRunits}{M_{\odot} \ \text{s}^{-1}}
\newcommand{\fbin}{f_{\text{bin}}}
\newcommand{\fact}{f_{\text{act}}}
\newcommand{\fsurv}{f_{\text{surv}}}

\newcommand{\JLW}{J_{\text{LW}}}

\newcommand{\fion}{f_{\text{ion}}}
\newcommand{\nnu}{$n_{\nu}$}
\newcommand{\ncol}{N_i}
\newcommand{\Tvir}{T_{\text{vir}}}

\newcommand{\NHI}{N_{H\textsc{i}}}

\newcommand{\emissivity}{\text{erg} \ \text{s}^{-1} \ \text{Hz}^{-1} \ \text{cMpc}^{-3}}

\newcommand{\Ja}{J_{\alpha}}
\newcommand{\Lya}{\text{Ly-}\alpha}
\newcommand{\Lyn}{\text{Ly-}n}
\newcommand{\TS}{T_{\text{S}}}
\newcommand{\Ts}{T_{\text{S}}}
\newcommand{\TK}{T_{\text{K}}}
\newcommand{\Tk}{T_{\text{K}}}
\newcommand{\Tcmb}{T_{\text{CMB}}}
\newcommand{\TCMB}{T_{\text{CMB}}}
\newcommand{\TR}{T_{\text{R}}}
\newcommand{\Tr}{T_{\text{R}}}

% Physical constants
\newcommand{\kB}{k_{\text{B}}}

\newcommand{\esc}{\text{esc}}

% Overlap regions
\newcommand{\IV}{\mathcal{V}}
\newcommand{\OV}{\mathcal{O}}
\newcommand{\pos}{\mathbf{x}}
\newcommand{\pospr}{\mathbf{x}^{\prime}}
\newcommand{\xpr}{x^{\prime}}

\newcommand{\fheat}{f^{\text{heat}}}
\newcommand{\fXh}{f_{X,h}}
\newcommand{\fioni}{f_i^{\text{ion}}}
\newcommand{\Lbol}{\mathcal{L}_{\text{bol}}}
\newcommand{\spec}{\mathcal{N}}
\newcommand{\Heat}{\mathcal{H}}
\newcommand{\trec}{$t_{\text{rec}}$}
\newcommand{\Lbox}{L_{\mathrm{box}}}
\newcommand{\dx}{\Delta x}
\newcommand{\dd}{\text{d}}

\newcommand{\MUV}{M_{\text{UV}}}

\newcommand{\drIF}{$\Delta r_{\mathrm{IF}}$}
\newcommand{\dTb}{$\delta T_b$}
\newcommand{\Nvec}{\mathbf{N}}
\newcommand{\sh}{\mathrm{sh}}
\newcommand{\Mdot}{\dot{M}}
\newcommand{\Ledd}{L_{\text{edd}}}
\newcommand{\intensityunits}{\text{erg} \ \text{s}^{-1} \ \text{cm}^{-2} \ \mathrm{Hz}^{-1} \ \text{sr}^{-1}}
\newcommand{\intensityunitsnumber}{\text{s}^{-1} \ \text{cm}^{-2} \ \mathrm{Hz}^{-1} \ \text{sr}^{-1}}
\newcommand{\sfrdunits}{M_{\odot} \ \text{yr}^{-1} \ \text{cMpc}^{-3}}

\chapter{Astrophysics from the 21-cm background}

\begin{bf}
  Jordan Mirocha (McGill University) \\
  %\author{Jordan Mirocha (McGill University)} \\
  
The goal of this chapter is to describe the astrophysics encoded by the 21-cm background. We will begin in \S\ref{sec:igm} with a brief introduction to the radiative transfer and ionization chemistry relevant to the high-$z$ intergalactic medium. Then, in \S\ref{sec:sources}, we will provide a review of the most plausible sources of ionization and heating in the early Universe. In \S\ref{sec:predictions}, we will explore the variety of current 21-cm predictions, and illustrate the dependencies of the global 21-cm signal and power spectrum to parameters of interest.
\end{bf}

%Figures 
%\begin{itemize}
%	\item Picture of reionization simulation.
%	\item Schematic of ray tracing
%	\item Show 1-D profiles to build intuition?
%	\item Show mean ionization and temperature histories from published work, defer on details of modeling assumptions to later sections.
%	\item Stellar spectra
%	\item XRB spectra 
%	\item Empirical constraints on $L_X$-SFR.
%\end{itemize}
%
%
%Here's my approach:
%\begin{itemize}
%	\item Talk about how 21-cm traces ionization and heating. Outline generic non-Eq chemistry setup and how one would do this in ``all its glory.''
%	\item Motivate separation of ionization and heating (mean free path), and how that allows more approximate techniques and useful conceptual framework. Outline those approximate techniques.
%	\item Turn to the sources. We've discussed how to model ionization and heating but not what the source terms are. Focus on evolution of individual sources and source populations (i.e., frequency bit then redshift/R bit)
%	\item Put it all together: basic predictions. Intuition for timing of different features in global signal and power spectrum, prospects for breaking degeneracies between different sources/parameters. Discussion of available tools, differences, progress? Lump in with previous section.
%\end{itemize}

%%%
%% Ionization, thermal, and Ly-a histories
%%%
\section{Properties of the High-$z$ Intergalactic Medium} \label{sec:igm}
In this section we provide a general introduction to the intergalactic medium (IGM) and how its properties are expected to evolve with time. We will start with a brief recap of the 21-cm brightness temperature (\ref{sec:dTb}), then turn our attention to its primary dependencies, the ionization state and temperature of the IGM (\S\ref{sec:chemistry}). In \S\ref{sec:smallscales}- \S\ref{sec:lya} we briefly review the radiative transfer (RT) relevant to modeling ionization, heating, and $\Lya$ coupling. Readers familiar with the basic physics may skip ahead to \S\ref{sec:sources}, in which we focus on the astrophysical sources most likely to heat and ionize the IGM at early times.

% T_21
\subsection{The brightness temperature} \label{sec:dTb}
The differential brightness temperature of a patch of the IGM at redshift $z$ and position $\mathbf{x}$ is given by\footnote{Refer back to Chapter 1 for a more detailed introduction.} 
\begin{equation}
    \delta T_b(z, \mathbf{x}) \simeq 27 (1 + \delta) (1 - x_i) \left(\frac{\Omega_{b,0} h^2}{0.023} \right) \left(\frac{0.15}{\Omega_{m,0} h^2} \frac{1 + z}{10} \right)^{1/2} \left(1 - \frac{\TR}{\TS} \right) , \label{eq:dTb}
\end{equation}
where $\delta$ is the baryonic overdensity relative to the cosmic mean, $x_i$ is the ionized fraction, $\TR$ is the radiation background temperature (generally the CMB, $\TR = \Tcmb$), and
\begin{equation}
    \TS^{-1} \approx \frac{\TR^{-1} + x_c \TK^{-1} + x_{\alpha} T_{\alpha}^{-1}}{1 + x_c + x_{\alpha}} . \label{eq:Ts}
\end{equation}
is the spin temperature, which quantifies the level populations in the ground state of the hydrogen atom, and itself depends on the kinetic temperature, $\TK$, and ``colour temperature'' of the Lyman-$\alpha$ radiation background, $T_{\alpha}$. Because the IGM is optically thick to Ly-$\alpha$ photons, the approximation $T_{\alpha} \approx \TK$ is generally very accurate.

The collisional coupling coefficients, $x_c$, themselves depend on the gas density, ionization state, and kinetic temperature (see \cite{Zygelman2005} for details). The radiative coupling coefficient, $x_{\alpha}$, depends on the Ly-$\alpha$ intensity, $J_{\alpha}$, via
\begin{equation}
    x_{\alpha} = \frac{S_{\alpha}}{1+z} \frac{J_{\alpha}}{{J}_{\alpha,0}} \label{eq:xalpha}
\end{equation}
where
\begin{equation}
    J_{\alpha,0} \equiv \frac{16\pi^2 T_{\star} e^2 f_{\alpha}}{27 A_{10} T_{\gamma,0} m_e c} . \label{eq:Jnorm}
\end{equation}
$J_{\alpha}$ is the angle-averaged intensity of Ly-$\alpha$ photons in
units of $\intensityunitsnumber$, $S_{\alpha}$ is a correction factor that
accounts for variations in the background intensity near line-center
\cite{Chen2004,FurlanettoPritchard2006,Hirata2006}, $m_e$ and $e$ are the
electron mass and charge, respectively, $f_{\alpha}$ is the $\Lya$ oscillator
strength, $T_{\gamma,0}$ is the CMB temperature today, and $A_{10}$ is the Einstein A coefficient for the 21-cm transition.

A more detailed introduction to collisional and radiative coupling can be found in Chapter 1. For the purposes of this chapter, the key takeaway from Equations \ref{eq:dTb}-\ref{eq:xalpha} is simply that the 21-cm background probes the ionization field, kinetic temperature field, and $\Lya$ background intensity. We quickly review the basics of non-equilibrium ionization chemistry in the next sub-section (\S\ref{sec:chemistry}) before moving on to sources of heating, ionization, and the $\Lya$ background in \S\ref{sec:sources}.

% Global signal
%\subsubsection{The ``global'' 21-cm signal}
%On very large scales...
%\begin{align}
%    \delta T_b \simeq 27 (1 - \mathbf{x_i}) \left(\frac{\Omega_{b,0} h^2}{0.023} \right) \left(\frac{0.15}{\Omega_{m,0} h^2} \frac{1 + z}{10} \right)^{1/2} \left(1 - \frac{T_R}{T_S} \right) , \label{eq:dTb}
%\end{align}
%
%Many experiments are targeting this signal. For this reason, modeling efforts for the global signal often take an approximate approach. Under the assumption that fluctuations in $\delta$, $x_i$, and $T_S$ are uncorrelated, the volume-averaged differential brightness temperature is simply related to the volume-averaged density, ionization fraction, and spin temperature. Averaging over large volumes means $\delta \approx 0$, and while in general these fields \textit{will} be correlated, {\color{red} the effects are likely minor: cite that one paper that Xueli Chen is on.}
%
%
%In the next three sections, we walk through the main epochs of evolution relevant to the 21-cm background, starting with reionization, and working our way backwards in time to first light. As in this section, boldfaced symbols refer to variables with an implicit spatial dependence, while regularly typset symbols refer to the spatial average. {\color{red} is this too tedious?}
%
%Talk here about how in numerical simulations we would just do radiative transfer so there's no need to break up all these things. But, RT is expensive, so in practice most models (at least those used for inference) make approximations, and it is very convenient to consider

%%
% RT background?
%%
\subsection{Basics of Non-Equilibrium Ionization Chemistry} \label{sec:chemistry}
As described in the previous section, the 21-cm brightness temperature of a patch of the IGM depends on the ionization and thermal state of the gas, as well as the incident Ly-$\alpha$ intensity\footnote{Note that Ly-$\alpha$ photons can transfer energy to the gas (see, e.g., \cite{Venumadhav2018}) though we omit this dependence from the current discussion (see Ch. 1.)}. The evolution of the ionization and temperature are coupled and so must be evolved self-consistently. The number density of hydrogen and helium ions in a static medium can be written as the following set of coupled differential equations:
\begin{align}
    \frac{d \nHII}{dt} & = (\ionHI + \ionsecHI + \ioncollHI \nel) \nHI - \recHII \nel \nHII   \label{eq:HIIRateEquation} \\
    \frac{d \nHeII}{dt} & = (\ionHeI + \ionsecHeI + \ioncollHeI \nel) \nHeI \nonumber + \recHeIII \nel \nHeIII  - (\ioncollHeII + \recHeII + \xiHeII) \nel \nHeII \\ & - (\ionHeII + \ionsecHeII) \nHeII \label{eq:HeIIRateEquation} \\ 
    \frac{d \nHeIII}{dt} & = (\ionHeII + \ionsecHeII + \ioncollHeII \nel) \nHeII  - \recHeIII \nel \nHeIII . \label{eq:HeIIIRateEquation} .
\end{align}
Each of these equations represents the balance between ionizations of species
\HI, \HeI, and \HeII, and recombinations of \HII, \HeII, and
\HeIII. Associating the index $i$ with absorbing species, $i = $\HI, \HeI,
\HeII, and the index $i^{\prime}$ with ions, $i^{\prime} = $\HII, \HeII,
\HeIII, we define $\Gamma_i$ as the photo-ionization rate coefficient,
$\gamma_i$ as the rate coefficient for ionization by photo-electrons \cite[; see Ch. 1]{Shull1985,Furlanetto2010}, $\alpha_{i^{\prime}}$
($\xi_{i^{\prime}}$) as the case-B (dielectric) recombination rate
coefficients, $\beta_i$ as the collisional ionization rate coefficients, and
$\nel = \nHII + \nHeII + 2\nHeIII$ as the number density of electrons.

While the coefficients $\alpha$, $\beta$, and $\xi$ only depend on the gas temperature, the photo- and secondary-ionization coefficients, $\Gamma$ and $\gamma$, depend on input from astrophysical sources (see \S\ref{sec:sources}).

The final equation necessary in a primordial chemical network is that governing the kinetic temperature evolution, which we can write as a sum of various heating and cooling processes, i.e.,
\begin{align}
    \frac{3}{2}\frac{d}{dt}\left(\frac{\kB \TK \ntot}{\mu}\right) & = \fheat  \sum_i n_i \Lambda_i - \sum_i \zeta_i \nel n_i - \sum_{i^{\prime}} \eta_{i^{\prime}} \nel n_{i^{\prime}} \nonumber \\ & - \sum_i \psi_i \nel n_i - \cooldielHeII \nel \nHeII \label{eq:TemperatureEvolution} .
\end{align}
Here, $\Lambda_i$ is the photo-electric heating rate coefficient (due to
electrons previously bound to species $i$), $\cooldielHeII$ is the dielectric
recombination cooling coefficient, and $\zeta_i$, $\eta_{i^{\prime}}$, and
$\psi_i$ are the collisional ionization, recombination, and collisional
excitation cooling coefficients, respectively, where primed indices
$i^{\prime}$ indicate ions $\HII$, $\HeII$, and $\HeIII$, and unprimed
indices $i$ indicate neutrals $\HI$, $\HeI$, and $\HeII$. The constants in
Equation (\ref{eq:TemperatureEvolution}) are the total number density of
baryons, $\ntot = n_\mathrm{H} + n_{\mathrm{He}} + \nel$, the mean molecular
weight, $\mu$, Boltzmann's constant, $\kB$, and the fraction of photo-electron energy deposited as heat, $\fheat$ (sometimes denoted $f_{\mathrm{abs}}$) \cite{Shull1985,Furlanetto2010}. Formulae to compute the values of
$\alpha_i$, $\beta_i$, $\xi_i$, $\zeta_i$, $\eta_{i^{\prime}}$, $\psi_i$, and
$\cooldielHeII$, are compiled in, e.g., \cite{Fukugita1994,Hui1997}. Terms involving helium become increasingly important in a medium irradiated by X-rays.

These equations do not yet explicitly take into account the cosmic expansion, which dilutes the density and adds an adiabatic cooling term to Eq. \ref{eq:TemperatureEvolution}, however these generalizations are straightforward to implement in practice. For the duration of this chapter we will operate within this simple chemical network, ignoring, e.g., molecular species like $\Htwo$ and $\mathrm{HD}$ whose cooling channels are important in primordial gases. Though an interesting topic in their own right, molecular processes reside in the ``subgrid'' component of most 21-cm models, given that they influence how, when, and where stars are able to form (see \S\ref{sec:sources}), but do not directly affect the bulk properties of the IGM on large scales to which 21-cm measurements are sensitive. 

%%
% POINT SOURCES
%%
\subsection{Ionization and Heating Around Point Sources} \label{sec:smallscales}
In order to build intuition for the progression of ionization and heating in the IGM it is instructive to consider the impact of a single point source of UV and X-ray photons on its surroundings. Many early works focused on such 1-D radiative transfer problems \cite{Zaroubi2007,Thomas2008}. In principle, this is the ideal way to simulation reionization -- iterating over all sources in a cosmological volume and for each one applying 1-D radiative transfer techniques over the surrounding $4\pi$ steradians. In practice, such approaches are computationally expensive, and while they provide detailed predictions \cite{OShea2015,Ocvirk2016,Gnedin2014}, more approximate techniques are required to survey the parameter space and perform inference (see Chapter 4).

In 1-D, the change in the intensity of a ray of photons, $I_{\nu}$, is a function of the path length, $s$, the emissivity of sources along the path, $j_{\nu}$, and the absorption coefficient, $\alpha_{\nu}$, 
\begin{equation}
	dI_{\nu} = j_{\nu} - \alpha_{\nu} I_{\nu} .
\end{equation}
If considering a point source, $j_{\nu} = 0$, we can integrate this radiative transfer equation (RTE) to obtain
\begin{equation}
	I_{\nu}(s) = I_{\nu,0} \exp\left[-\int_0^s \alpha_{\nu}(s^{\prime}) ds^{\prime} \right] ,
\end{equation}
i.e., the intensity of photons declines exponentially along the ray. It is customary to define the optical depth, 
\begin{equation}
	d\tau_{\nu} \equiv \alpha_{\nu} ds ,
\end{equation}
in which case we can write
\begin{equation}
	I_{\nu}(s) = I_{\nu,0} e^{-\tau_{\nu}} .
\end{equation}
In the reionization context, the optical depth of interest is that of the IGM, which is composed of (almost) entirely hydrogen and helium\footnote{Note that there will be small-scale absorption as well, though in most models this is unresolved, and parameters governing photon escape are used to quantify this additional opacity (see \S\ref{sec:fesc} and \S\ref{sec:xray_esc}).}, in which case the optical depth is 
\begin{equation}
	\tau_{\nu} = \sum_i \sigma_{\nu,i} N_i
\end{equation}
where $i=\HI,\HeI,\HeII$, and $N_i = \int_0^s ds^{\prime} n_i(s^{\prime})$ is the column density of each species along the ray.

With a solution for $I_{\nu}(s)$ in hand, one can determine the photoionization and heating rates by integrating over all photon frequencies and weighting by the bound-free absorption cross section for each species. For example, the photoionization rate coefficient for hydrogen can be written as 
\begin{equation}
	\Gamma_{\HI}(s) = \int_{\nu_{\HI}}^{\infty} \sigma_{\HI} I_{\nu}(s) \frac{d\nu}{h\nu} \label{eq:GammaHI}
\end{equation}
where $\nu_{\HI}$ is the frequency of the hydrogen ionization threshold, $h\nu=13.6$ eV.

Note that in practice the RTE is solved on a grid, in which case it may be difficult to achieve high enough spatial resolution to ensure photon conservation. For example, a discretized version of Eq. \ref{eq:GammaHI} implies that the intensity of radiation incident upon the face of a resolution element determines the photoionization rate within that element. However, the radiation incident on the subsequent resolution element is not guaranteed to correctly reflect the attenuation within the preceding element. As a result, in order to guarantee photon conservation, it is common to slightly reframe the calculation as follows \cite{Abel1999}:
\begin{align}
    \Gamma_i & = A_i \int_{\nu_i}^{\infty} I_{\nu} e^{-\tau_{\nu}} \left(1 - e^{-\Delta \tau_{i,\nu}}\right) \frac{d\nu}{h\nu} \label{eq:PhotoIonizationRate} \\
    \gamma_{ij} & = A_j \int_{\nu_j}^{\infty} \left(\frac{\nu - \nu_j}{\nu_i}\right) I_{\nu} e^{-\tau_{\nu}} \left(1 - e^{-\Delta \tau_{j,\nu}}\right) \frac{d\nu}{h\nu} \label{eq:SecondaryIonizationRate} \\
    \Lambda_i & = A_i \int_{\nu_i}^{\infty} (\nu - \nu_i) I_{\nu} e^{-\tau_{\nu}} \left(1 - e^{-\Delta \tau_{i,\nu}}\right) \frac{d\nu}{\nu}  \label{eq:HeatingRate} .
\end{align}
The normalization constant in each expression is defined as $A_i \equiv L_{\mathrm{bol}}/n_i V_{\sh}(r)$, where $V_{\sh}$ is the volume of a shell in this 1-D grid of concentric spherical shells, each having thickness $\Delta r$ and volume $V_{\sh}(r) = 4 \pi [(r + \Delta r)^3 - r^3] / 3$, where $r$ is the distance between the origin and the inner interface of each shell. We denote the ionization threshold energy for species $i$ as $h\nu_i$. $I_{\nu}$ represents the SED of radiation sources, and satisfies $\int_{\nu} I_{\nu} d\nu = 1$, such that $L_{\mathrm{bol}} I_{\nu} = L_{\nu}$. Note that the total secondary ionization rate for a given species is the sum of ionizations due to the secondary electrons from all species, i.e., $\gamma_i = \fion \sum_j \gamma_{ij} n_j / n_i$.

These expressions preserve photon number by inferring the number of photo-ionizations of species $i$ in a shell from the radiation incident upon it and its optical depth \cite{Abel1999},
\begin{equation}
    \Delta \tau_{i,\nu} = n_i \sigma_{i,\nu} \Delta r .
\end{equation}    
This quantity is not to be confused with the total optical depth between source and shell, $\tau_{\nu} = \tau_{\nu}(r)$, which sets the incident radiation field upon each shell, i.e.,
\begin{align}
    \tau_{\nu}(r) & = \sum_i \int_0^r \sigma_{i,\nu} n_i(r^{\prime}) dr^{\prime} \nonumber \\
                  & = \sum_i \sigma_{i,\nu} \ncol(r) \label{eq:OpticalDepth}
\end{align}
where $\ncol$ is the column density of species $i$ at distance $r$ from the
source.

In words, Equations \ref{eq:PhotoIonizationRate}-\ref{eq:HeatingRate} are propagating photons from a source at the origin, with bolometric luminosity $L_{\mathrm{bol}}$, and tracking the attenuation suffered between the source and some volume element of interest at radius $r$, $e^{-\tau}$, and the attenuation within that volume element, $\Delta \tau$, which results in ionization and heating. In each case, we integrate over the contribution from photons at all frequencies above the ionization threshold, additionally modifying the integrands for $\gamma_{ij}$ and $\Lambda_i$ with $(\nu - \nu_i)$-like factors to account for the fact that both the number of photo-electrons (proportional to $(\nu - \nu_j) / \nu_i$) and their energy (proportional to $\nu - \nu_i$) determine the extent of secondary ionization and photo-electric heating. Equations \ref{eq:PhotoIonizationRate}-\ref{eq:HeatingRate} can be solved once a source luminosity, $L_{\mathrm{bol}}$, spectral shape, $I_{\nu}$, and density profile of the surrounding medium, $n(r)$, have been specified\footnote{In practice, to avoid performing these integrals on each step of an ODE solver (for Eqs. \ref{eq:HIIRateEquation}-\ref{eq:TemperatureEvolution}), the results can be tabulated as a function of $\tau$ or column density, $N_i$, where $\tau_{i,\nu}=\sigma_{i,\nu} N_i$ \cite{Thomas2008,Mirocha2012,Knevitt2014}.}.  

%\begin{figure*}[]
%\begin{center}
%\includegraphics[width=0.98\textwidth]{Mirocha/mcquinn2012_fig5.pdf}
%\end{center}
%\caption{{\bf Temperature profile around power-law sources in an expanding IGM \cite{McQuinn2012}. \textit{Right:} \cite{Knevitt2014}}}
%\label{fig:fzh04}
%\end{figure*}

\begin{figure*}[]
\begin{center}
\includegraphics[width=0.8\textwidth]{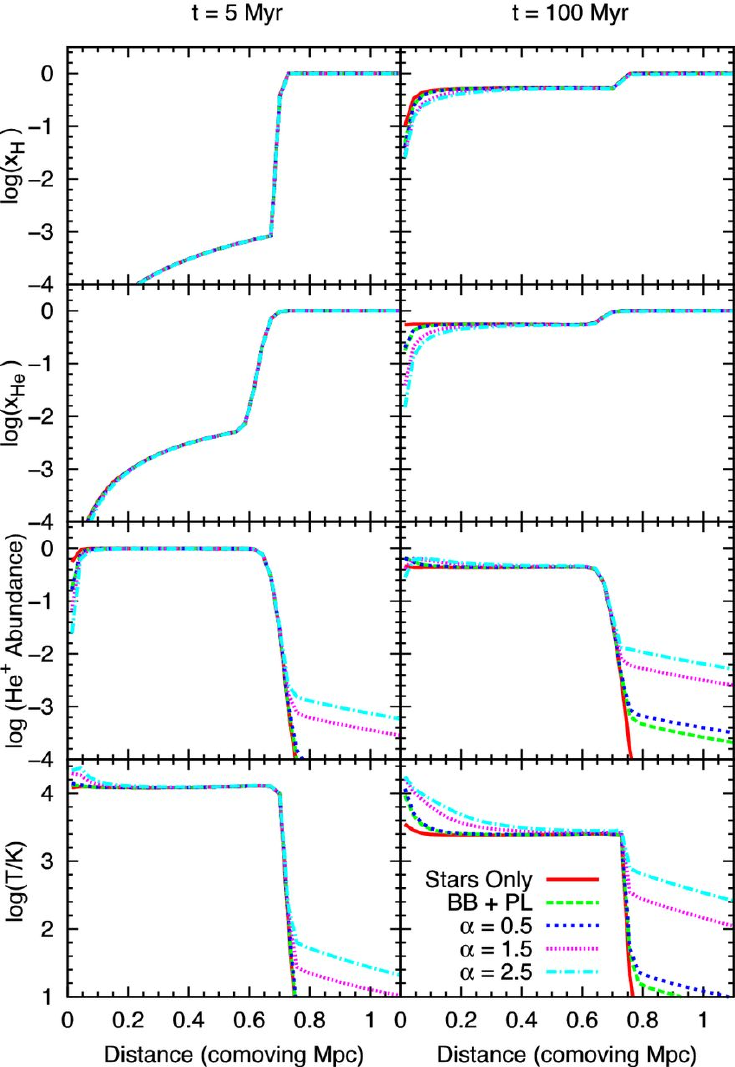}
\end{center}
\caption{{\bf Ionization and temperature profile around stellar UV and X-ray sources in a one-dimensional radiative transfer model \cite{Knevitt2014}.} From top to bottom, this includes the hydrogen neutral fraction, neutral helium fraction, singly-ionized helium fraction, and kinetic temperature, while the left and right columns indicate different time snapshots after sources are first turned on. Different lines adopt different source models, from a ``stars only'' model (solid red), to hybrid models with stars and X-ray sources with different power-law spectra (dashed and dotted curves). }
\label{fig:rt1d}
\end{figure*}

Figure \ref{fig:rt1d} shows an example 1-D radiative transfer model including sources of UV and X-ray photons. Because the mean free paths of UV photons are short, they generate sharp features in radial profiles of the neutral fraction of hydrogen (top) and helium (second row). The addition of X-ray sources largely influences the abundance of singly-ionized helium (third row) and the extended temperature structure beyond the fully-ionized region (fourth row) without dramatically modifying the sharp structures in of the hydrogen and helium fraction. 

Calculations like those shown in Figure \ref{fig:rt1d} motivate two-phase models of the IGM (e.g., \cite{Furlanetto2006,Pritchard2010,Mirocha2015}), in which UV photons carve out relatively distinct regions of fully-ionized hydrogen gas, while the hydrogen beyond these bubbles remains largely neutral\footnote{In practice one then solves two sets of equations like Eqs. \ref{eq:HIIRateEquation}-\ref{eq:TemperatureEvolution} -- one for each phase of the IGM. In the fully-ionized phase, the ionized fraction represents a volume-filling fraction, while in the ``bulk IGM'' phase, it retains its usual meaning.}. The mostly neutral ``bulk IGM'' outside of bubbles is affected predominantly by X-rays, which have mean free paths long enough to escape the environments in which they are generated (though see \S\ref{sec:xray_esc}). This also implies that the properties of a small patch of HI gas in the bulk IGM may be affected by many sources at cosmological distances. We focus on this limit in the next sub-section.

%%
% Metagalactic background
%%
\subsection{Ionization and Heating on Large Scales} \label{sec:largescales}
While the procedure outlined in the previous section is relevant to small-scale ionization and heating, i.e., that which is driven a single (or perhaps a few) source(s) close to a volume element of interest, it is also instructive to consider the ionization and heating caused by a \textit{population} of sources separated by great distances. In this limit, rather than considering the luminosity of a single source at the origin of a 1-D grid, we treat the volume-averaged emissivity of sources, $\epsilon_{\nu}$, in a large ``chunk'' of the Universe, and solve for the evolution of the mean intensity in this volume, $J_{\nu}$.

The transfer equation now takes its cosmological form, i.e., 
\begin{equation}
    \left(\frac{\partial}{\partial t} - \nu H(z) \frac{\partial}{\partial \nu} \right) J_{\nu}(z) + 3 H(z) J_{\nu}(z) =  \frac{c}{4\pi} \epsilon_{\nu}(z) (1 + z)^3 - c \alpha_{\nu} J_{\nu}(z) \label{eq:rte_diffeq}
\end{equation}
where $J_{\nu}$ is the mean intensity in units of $\intensityunits$, $\nu$ is the observed frequency of a photon at redshift $z$, related to the emission frequency, $\nu^{\prime}$, of a photon emitted at redshift $z^{\prime}$ as
\begin{equation}
    \nu^{\prime} = \nu \left(\frac{1 + z^{\prime}}{1 + z}\right) , \label{eq:EmissionFrequency}
\end{equation}
$\alpha_{\nu} = n \sigma_{\nu}$ is the absorption coefficient, not to be confused with recombination rate coefficient, $\alpha_{\HII}$, and $\epsilon_{\nu}$ is the co-moving emissivity of sources.

The optical depth, $d\tau = \alpha_{\nu} ds$, experienced by a photon at redshift $z$ and emitted at $z^{\prime}$ is an integral along a cosmological line element, summed over all absorbing species\footnote{In general, one must iteratively solve for $\overline{\tau}_{\nu}$ and $J_{\nu}$. However, in many models the bulk of cosmic re-heating precedes reionization, in which case $\overline{\tau}_{\nu}$ can be tabulated assuming a fully neutral IGM. This approach provides a considerable speed-up computationally and remains accurate even when reionization and reheating partially overlap \cite{Mirocha2014}.
}, i.e., 
\begin{equation}
    \overline{\tau}_{\nu}(z, z^{\prime}) = \sum_j \int_{z}^{z^{\prime}} n_j(z^{\dprime}) \sigma_{j, \nu^{\dprime}} \frac{dl}{dz^{\dprime}}dz^{\dprime} \label{eq:tau_igm}
\end{equation}

The solution to Equation \ref{eq:rte_diffeq} is
\begin{equation}
    \hat{J}_{\nu} (z) = \frac{c}{4\pi} (1 + z)^2 \int_{z}^{z_f} \frac{\epsilon_{\nu}^{\prime}(z^{\prime})}{H(z^{\prime})} e^{-\overline{\tau}_{\nu}} dz^{\prime} . \label{eq:AngleAveragedFlux}
\end{equation}    
where $z_f$ is the ``first light redshift'' when astrophysical sources first turn on, $H$ is the Hubble parameter, and the other variables take on their usual meaning\footnote{This equation can be solved efficiently on a logarithmic grid in $x\equiv 1+z$ \cite{Haardt1996,Mirocha2014}, in which case photons redshift seamlessly between frequency bins over time.}. 

With the background intensity in hand, one can compute the rate coefficients for ionization and heating. These coefficients are equivalent to those for the 1-D problem (Eqs. \ref{eq:PhotoIonizationRate}-\ref{eq:HeatingRate}), though the intensity of radiation at some distance $R$ from the source has been replaced by the mean background intensity,
\begin{align}
    \Gamma_{i}(z) & = 4 \pi n_i(z) \int_{\nu_{\min}}^{\nu_{\max}} \hat{J}_{\nu} \sigma_{\nu,i} d\nu  \\
    \gamma_{ij}(z) & = 4 \pi \sum_j n_j \int_{\nu_{\min}}^{\nu_{\max}}  \hat{J}_{\nu} \sigma_{\nu,j} (h\nu - h\nu_j) \frac{d\nu}{h\nu}  \\
    \epsilon_X(z) & = 4 \pi \sum_j n_j \int_{\nu_{\min}}^{\nu_{\max}}  \hat{J}_{\nu}  \sigma_{\nu,j} (h\nu - h\nu_j) d\nu
\end{align}
Then, the ionization state and temperature of the gas can be updated accordingly via Equations \ref{eq:HIIRateEquation}-\ref{eq:TemperatureEvolution}.

\begin{figure*}[]
\begin{center}
\vspace{50pt} \includegraphics[width=0.33\textwidth]{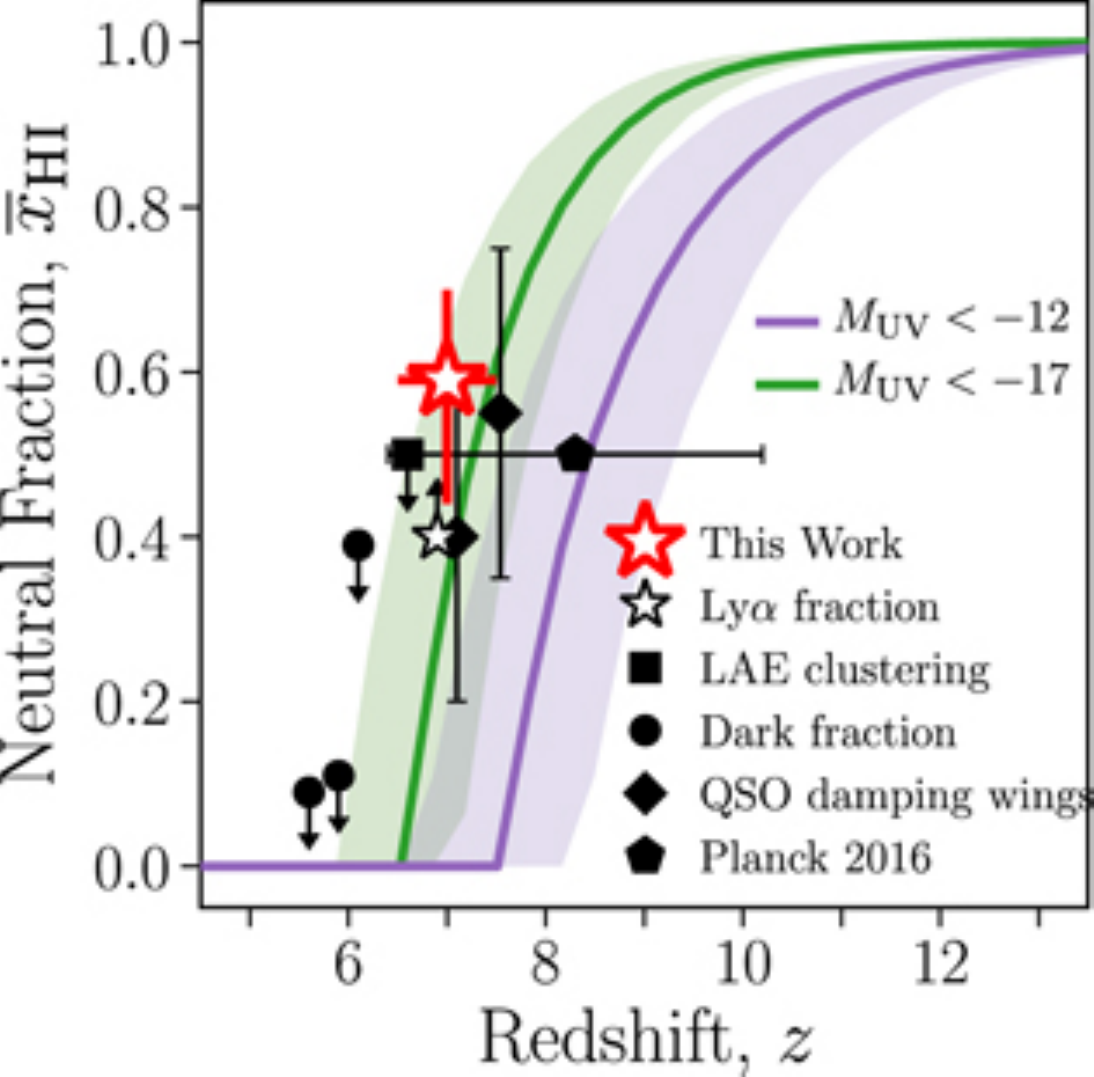} \hspace{20pt} 
\includegraphics[width=0.43\textwidth]{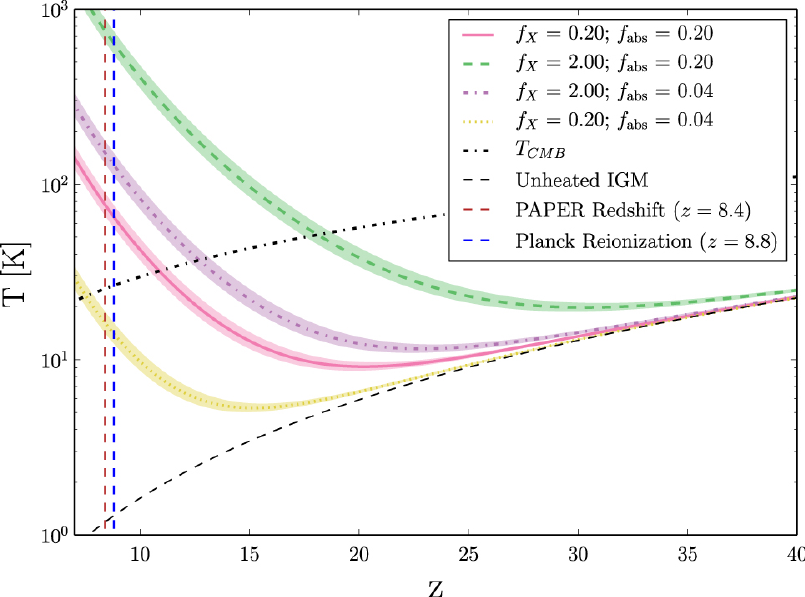}
\end{center}
\caption{{\bf Predictions for the evolution of the mean properties of the IGM.} \textit{Left:} Predictions for the mean neutral fraction of the IGM as a function of redshift compared to several observational constraints \cite{Mason2018}. Magenta curve includes all galaxies brighter than UV magnitude $\MUV < -12$, while green curve includes only brighter galaxies with $\MUV < -17$. \textit{Right:} Predictions for the mean kinetic temperature of the IGM \cite{Pober2015} for different assumptions about how efficiently galaxies produce X-rays (parameterized via $f_X$ and the fraction of X-ray energy absorbed in the IGM, $f_{\mathrm{abs}}$; see \S\ref{sec:sources}), compared to an early power spectrum limit from \textsc{paper} \cite{Parsons2014,Ali2015}. Note that the \textsc{paper} limit has since been revised \cite{Cheng2018,Kolopanis2019}.}
\label{fig:mean_igm}
\end{figure*}

Figure \ref{fig:mean_igm} shows predictions for the evolution of the mean ionized fraction and kinetic temperature of the IGM \cite{Mason2018,Pober2015} using the two-phase IGM picture described above. While current observations are consistent with reionization occurring relatively rapidly at $z \lesssim 10$, heating is generally more gradual, and so far unconstrained. The strongest 21-cm signals occur when the IGM remains cold during reionization, so upper limits on the amplitude of 21-cm signals translate to lower limits on the efficiency of X-ray heating in the early Universe.

%%
% WF COUPLING
%%
\subsection{Ly-$\alpha$ Coupling} \label{sec:lya}
On scales large and small, the 21-cm background will only probe the kinetic temperature of the gas if the $\Lya$ background intensity is strong enough to couple the spin temperature to the kinetic temperature. Determining the $\Lya$ background intensity, $J_{\alpha}$, requires a special solution to the cosmological radiative transfer equation (Eq. \ref{eq:rte_diffeq}). Two effects separate this problem from the generic transfer problem outlined in \S\ref{sec:largescales}: (i) the Lyman series forms a set of horizons for photons in the $10.2 < h \nu / \mathrm{eV} < 13.6$ interval, giving rise to the so-called ``sawtooth modulation'' of the soft UV background \cite{Haiman1997}, and (ii) the Ly-$\alpha$ background is sourced both by photons redshifting into the line resonance as well as those produced in cascades downward from higher $n$ transitions \cite{Pritchard2006}.

As a result, it is customary to solve the RTE in each $\Lyn$ frequency interval separately. Within each interval, bounded by Ly-$n$ line on its red edge and Ly-$n+1$ on its blue edge, the optical depth is small in a primordial medium because no photon redward of the Lyman edge can ionize hydrogen or helium\footnote{There is in principle a small opacity contribution from $H_2$, though we neglect this in what follows as the $H_2$ fraction in the IGM is expected to be small.}. As a result, any photon starting its journey just redward of Ly-$\beta$ will travel freely until it redshifts into the Ly-$\alpha$ resonance, while photons originating at bluer wavelengths will encounter Ly-$n$ resonances (with $n>2$), only a fraction of which will ultimately result in $\Lya$ photons.

We can thus write the mean $\Lya$ background intensity as
\begin{equation}
    \widehat{J}_{\alpha}(z) = \frac{c}{4\pi} (1 + z)^2 \sum_{n = 2}^{\nmax} \frecn \int_z^{z_{\max}^{(n)}} \frac{\epsilon_{\nu}^{\prime}(z^{\prime})}{H(z^{\prime})} dz^{\prime} \label{eq:LymanAlphaFlux}
\end{equation}
where $\frecn$ is the ``recycling fraction,'' that is, the fraction of photons that redshift into a Ly-$n$ resonance that ultimately cascade through the $\Lya$ resonance \cite{Pritchard2006}. The upper bound of the definite integral,
\begin{equation}
    1 + z_{\max}^{(n)} = (1 + z) \frac{\left[1 - (n + 1)^{-2}\right]}{1 - n^{-2}} ,
\end{equation}
is set by the horizon of $\Lyn$ photons -- a photon redshifting through the  $\Lyn$ resonance at $z$ could only have been emitted at $z^{\prime} < z_{\max}^{(n)}$, since emission at slightly higher redshift would mean the photon redshifted through the $\text{Ly}(n+1)$ resonance. The sum over Ly-$n$ levels in Eq. \ref{eq:LymanAlphaFlux} is generally truncated at $n_{\max}=23$ \cite{Barkana2005} since the horizon for such photons is smaller than the typical ionized bubble sourced by an individual galaxy. As a result, any $\Lya$ photons generated by such high-$n$ cascades are ``wasted'' as far as the spin temperature is concerned, as they will most likely have redshifted out of resonance before reaching any neutral gas. $\Lya$ emission produced by recombinations in galactic HII regions is generally neglected for the same reason. Though there are some assumptions built into the $n_{\max}$ estimate, the total $\Lya$ photon budget is relatively insensitive to the exact value of $n_{\max}$ \cite{Barkana2005,Pritchard2006}.

Note that in general the mean free path of photons between Lyman series resonances is very long, which makes tracking them in numerical simulations  very expensive. For example, a photon emitted just redward of Ly-$\beta$ and observed at the Ly-$\alpha$ frequency at redshift $z$ has traveled a distance
\begin{equation}
	d_{\beta\rightarrow \alpha} \simeq 200 \ h_{70}^{-1} \left(\frac{\Omega_{m,0}}{0.3} \frac{1+z}{20} \right)^{-1/2} \ \mathrm{cMpc},
\end{equation}
where we have assumed the high-$z$ approximation $\Omega_{\Lambda} \ll \Omega_m$. This exceeds a Hubble length at high-$z$, meaning most of the Wouthuysen-Field coupling at very early times must come from photons originating just blueward of their nearest $\Lyn$ resonance. Despite their long mean free paths, fluctuations in the $\Lya$ background inevitably arise \cite{Barkana2005,Ahn2009,Holzbauer2012}. However, this background is expected to become uniform (and strong) relatively quickly, meaning in general the 21-cm background is only sensitive to $J_{\alpha}$ at the earliest epochs (see \S\ref{sec:dep_alpha}).

\section{Sources of the UV and X-ray Background} \label{sec:sources}
In the previous section we outlined the basic equations governing the ionization and temperature evolution of the IGM without actually specificying the sources of ionization and heating\footnote{Note that some (at least roughly) model-independent constaints on the properties of the IGM should be attainable with future 21-cm measurements \cite{Cohen2017,Cohen2018,Mirocha2013}.}. Instead, we used a placeholder emissivity, $\epsilon_{\nu}$, to encode the integrated emissions of sources at frequency $\nu$ within some region $R$. We will now write this emissivity as an integral over the differential luminosity function (LF) of sources, $dn/dL_{\nu}$, i.e.,
\begin{equation}
	\epsilon_{\nu}(z,R) = \int_0^{\infty} dL_{\nu} \frac{dn}{dL_{\nu}} .
\end{equation}
where $\nu$ refers to the rest frequency of emission at redshift $z$. 

It is common to rewrite the emissivity as an integral over the DM halo mass function (HMF), $dn/dm$, multiplied by a conversion factor between halo mass and galaxy light, $dm/dL_{\nu}$, i.e.,
\begin{equation}
	\epsilon_{\nu}(z,R) = \int_{\mmin}^{\infty} dm \frac{dn}{dm} \frac{dm}{d L_{\nu}} ,
\end{equation}
where $\mmin$ is the minimum mass of DM halos capable of hosting galaxies. Because $dn/dm$ is reasonably well-determined from large N-body simulations of structure formation \cite{PS1974,SMT2001,Tinker2010}, much of the modeling focus is on the mass-to-light ratio, $dm/dL_{\nu}$, which encodes the efficiency with which galaxies form in halos and the relative luminosities of different kinds of sources within galaxies (e.g., stars, compact objects, diffuse gas) that emit at different frequencies\footnote{Most models consider regions $R$ that are sufficiently large that one can assume a well-populated HMF, though at very early times this approximation may break down, rendering stochasticity due to poor HMF sampling an important effect.}. 

The main strength of the 21-cm background as a probe of high-$z$ galaxies is now apparent: though 21-cm measurements cannot constrain the properties of individual galaxies, they can constrain the properties of \textit{all} galaxies, in aggregate, \textit{even those too faint to be detected directly}. As a result, it is common to forego detailed modeling of the mass-to-light ratio and instead relate the emissivity to the fraction of mass in the Universe in collapsed halos,
\begin{equation}
	\epsilon_{\nu}(z, R) = \rho_b \fcoll(z, R) \zeta_{\nu} ,
\end{equation}
where $\rho_b$ is the baryon mass density, the collapsed fraction is an integral over the HMF,
\begin{equation}
	\fcoll = \rho_m^{-1} \int_{\mmin}^{\infty} dm m \frac{dn}{dm}
\end{equation}
and $\zeta_{\nu}$ is an efficiency factor that quantifies the number of photons emitted at frequency $\nu$ per baryon of collapsed mass in the Universe. It is generally modeled as
\begin{equation}
	\zeta_{\nu} = f_{\ast} N_{\nu} f_{\esc,\nu} , \label{eq:zeta}
\end{equation}
where $f_{\ast}$ is the star formation efficiency (SFE), in this case defined to be the fraction of baryons that form stars, $N_{\nu}$ is the number of photons emitted per stellar baryon at some frequency $\nu$, and $f_{\esc,\nu}$ is the fraction of those photons that escape into the IGM. One could define additional $\zeta$ factors to represent, e.g., emission from black holes or exotic particles, in which case $f_{\ast}$ and $N_{\nu}$ would be replaced by some black hole or exotic particle production efficiencies. In practice, most often three $\zeta$ factors are defined: $\zeta=\zeta_{\mathrm{ion}}$, $\zeta_X$, and $\zeta_{\alpha}$, i.e., one efficiency factor for each radiation background that influences the 21-cm signal. A minimal model for the 21-cm background thus contains four parameters: $\mmin$, $\zeta$, $\zeta_X$, and $\zeta_{\alpha}$. Note that  $\zeta_X$ and $\zeta_{\alpha}$ are often replaced by the parameters $f_X$ and $f_{\alpha}$, where the latter are defined such that $f_X=1$ and $f_{\alpha}=1$ correspond to fiducial values of $\zeta_X$ and $\zeta_{\alpha}$.

Because the factors constituting $\zeta$ are degenerate with each other, at least as far as 21-cm measurements are concerned, they generally are not treated separately as free parameters. However, it is still useful to consider each individually in order to determine a fiducial value of $\zeta$ and explore deviations from that fiducial model. In addition, inclusion of ancillary measurements may eventually allow $\zeta$ to be decomposed into its constituent parts \cite{Mirocha2017,Park2019,Greig2019}. For the remainder of this section, we focus on plausible values of $f_{\ast}$, $N_{\nu}$ and $f_{\esc,\nu}$, and the extent to which these quantities are currently understood.

%%
% PopII Star Formation
%%
\subsection{Star Formation} \label{sec:sfe}
Though a first-principles understanding of star formation remains elusive, the bulk properties of the star-forming galaxy population appear to obey simple scaling relationships. In this section, we outline the basic strategies used to infer the relationships between star formation and dark matter halos, and how such relationships can be used to inform 21-cm models.

The simplest description of the galaxy population follows from the assumption that each dark matter halo hosts a single galaxy. With a model for the abundance of DM halos, i.e., the halo mass function (HMF), many of which are readily available \cite{PS1974,SMT2001}, one can then ``abundance match'' halos with measured galaxy abundances \cite{Bouwens2015,Finkelstein2015}, i.e.,
\begin{align}
	n(>L_h) & = \int_L^{\infty} \frac{dn}{dL^{\prime}} dL^{\prime} \nonumber \\
	& = n(>m_h)  \nonumber \\
	& = \int_{m_h}^{\infty} \frac{dn}{dm_h^{\prime}} dm_h^{\prime} .
\end{align}
This procedure reveals the mapping between mass and light, $dL/dm_h$, upon repeated integration over a grid of $L_h$ values, solving for the $M_h$ value needed for abundances to match. 

Results of this simple procedure show that galaxy luminosity is a function of both halo mass and cosmic time \cite{Trenti2009,Moster2010,Behroozi2013,Tacchella2018,Mashian2016,Sun2016,Mason2015}. While the HMF can be readily used to predict the abundances of halos out to arbitrary redshifts, one shortcoming of this approach is that any evolution in $dL/dM$ must be modeled via extrapolation. As a result, predictions for deeper and/or higher redshift galaxy surveys may not be physically motivated.

To avoid the possibility of unphysical extrapolations of $dL/dm_h$, it is becoming more common to parameterize the galaxy -- halo connection from the outset, effectively resulting in forward models for galaxy formation that link galaxy star formation rate (SFR), $\dot{m}_{\ast}$, to halo mass, $m_h$, or mass accretion rate (MAR), $\dot{m}_h$, e.g.,
\begin{equation}
	\dot{m}_{\ast}(z,m_h) = \tilde{f}_{\ast}(z,m_h) m_h (z,m_h) . \label{eq:SFE_M}
\end{equation}
or
\begin{equation}
	\dot{m}_{\ast}(z,m_h) = f_{\ast}(z,m_h) \dot{m}_h (z,m_h) . \label{eq:SFE_MAR}
\end{equation}
The star formation efficiency (SFE), here indicated with $\tilde{f}_{\ast}$ and $f_{\ast}$, to explicitly indicate whether tied to $m_h$ or $\dot{m}_h$, is left as a flexible function to be calibrated empirically\footnote{Note that $\tilde{f}_{\ast}$ in Equation \ref{eq:SFE_M} necessarily has units of $\mathrm{time}^{-1}$, whereas $f_{\ast}$ in Eq. \ref{eq:SFE_MAR} is dimensionless.}. In the MAR-based model, one of course requires a model for the halo MAR as well as the HMF, though such models are readily available from the results of numerical simulations \cite{McBride2009,Trac2015}, or modeled approximately from the HMF itself \cite{Furlanetto2017}. Both approaches are used in the literature, and while inferred SFRs are largely in agreement, there is some difference in the interpretation of the models, which we revisit below in \S\ref{sec:sfe_physical}.

Finally, to complete the link between halos and galaxies, one must adopt a conversion factor between SFR and galaxy luminosity in some band. High-$z$ measurements mostly probe the rest UV spectrum of galaxies, so it is customary to link the SFR with the rest $1600 \ \AA$ luminosity of galaxies,
\begin{equation}
	L_{1600}(z, m_h) = l_{1600} \dot{m}_{\ast}(z,m_h) \label{eq:L1600}
\end{equation}
where $l_{1600}$ is of order $10^{28} \ \mathrm{erg} \ \mathrm{s}^{-1} \ \mathrm{Hz}^{-1} \ (M_{\odot} / \mathrm{yr})^{-1}$ according to commonly-used stellar population synthesis models, assuming constant star formation \cite{Leitherer1999,Eldridge2009,Conroy2009}. The precise value depends on stellar metallicity, binarity, and initial mass function (IMF), and varies from model to model. We will revisit the details of stellar spectra in \S\ref{sec:UV}, as there is a clear degeneracy between the assumed UV properties of galaxies and the inferred SFE.

\begin{figure*}[]
\begin{center}
\includegraphics[width=0.98\textwidth]{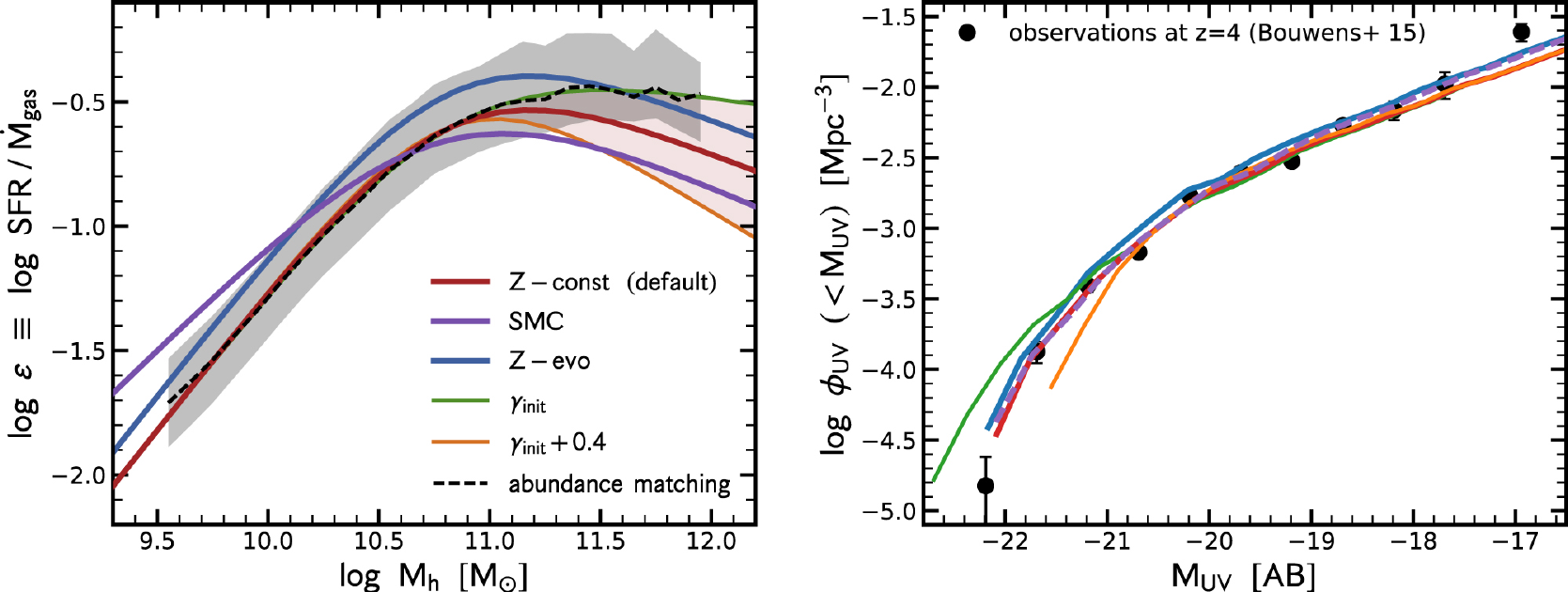}
\end{center}
\caption{{\bf Relationship between halos and star formation recovered via semi-empirical modeling \cite{Tacchella2018}.} \textit{Left:} Star formation efficiency as a function of halo mass for a variety of different approaches, including constant metallicity ``Z-const'', an evolving metallicity model ``Z-evo'', a model with SMC dust instead of the standard relation from \cite{Meurer1999} (see \S\ref{sec:dust}), and the pure abundance matching approach. \textit{Right:} Modeled luminosity function at $z=4$ compared to measurements from \cite{Bouwens2015}.}
\label{fig:sfe_lf}
\end{figure*}

The end result of this exercise is a calibrated SFE curve, which can then be used to make predictions for galaxy properties too faint or too distant to have been detected by current surveys. A representative example \cite{Tacchella2018} is shown in Figure \ref{fig:sfe_lf} (which includes a common dust correction -- see \S\ref{sec:dust}). The rise and fall in the SFE is a generic result of semi-empirical models \cite{Mason2015,Sun2016,Mashian2016,Tacchella2018,Behroozi2019}, indicating a change in how galaxies form stars in halos above and below $\sim 10^{12} \ M_{\odot}$. Such models generally agree that star formation is inefficient, with peak values $f_{\ast} \lesssim 0.1$, and $f_{\ast} \lesssim 0.01$ in the less massive (but numerous) population of galaxies residing in halos $M_h \lesssim 10^{11} \ M_{\odot}$. As a result, if using the standard $\zeta$ modeling approach (see Eq. \ref{eq:zeta}), reasonable fiducial $f_{\ast}$ values are $f_{\ast} \sim 0.01-0.1$. Of course, evolution of the HMF implies that representative values of $f_{\ast}$ will also evolve with time, though discerning such effects will only be possible in 21-cm analyses focused on a broad frequency range. 

%%
% Does this make sense?
%%
\subsubsection{Physical arguments for inferred behavior of $f_{\ast}$} \label{sec:sfe_physical}
Current high-$z$ measurements support a relatively simple picture of star formation in early galaxies, in which galaxies maintain a rough equilibrium between inflow and outflow through stellar feedback \cite{Bouche2010,Dave2012,Dekel2014}, the efficiency of which is a strong function of halo mass and perhaps time. Though there are quantitative differences amongst studies in the literature, which could arise due to different assumptions about dust, stellar populations, and/or different definitions of $f_{\ast}$, there does appear to be a consensus that star formation is maximally efficient (and feedback correspondingly inefficient) in $M_h \simeq 10^{11.5}-10^{12} M_{\odot}$ halos \cite{Mason2015,Sun2016,Moster2010,Mashian2016,Tacchella2018,Behroozi2019}. The decline in the SFE below the peak is widely thought to be a signature of stellar feedback, while the decline in massive systems is likely due to shock heating and/or AGN feedback, which reduces the availability of cold gas  \cite{FG2011,Croton2006,Somerville2008}. 

This general trend can be explained -- but perhaps not understood -- from relatively simple arguments. The basic idea is that star formation is fueled by the inflow of gas from the IGM, but that the overall rate of star formation in galaxies is self-regulated by feedback from stellar winds and supernovae explosions, both of which expell gas that could otherwise form stars \cite{Dayal2014,Furlanetto2017}. As a result, galaxies forming in shallow gravitational potential wells are at a disadvantage simply because the escape velocity is lower, making it easier for supernovae and winds to drive material out of the galaxy. However, the escape velocity depends on both the mass and size of an object -- if low-mass halos are sufficiently compact, they may be able to retain enough gas to continue forming stars.

To build some intuition for possible outcomes, it is common to model star formation as a balance between inflow and outflow \cite{Bouche2010,Dave2012,Dekel2014}, i.e.,
\begin{equation}
	\dot{m}_{\ast} = \dot{m}_b - \dot{m}_w
\end{equation}
where $m_b$ is the accretion rate of baryons onto a halo and $\dot{m}_w$ is the mass-loss rate through winds (and/or supernovae). If we relate mass-loss to star formation via ``mass loading factor'' $\eta$, $\dot{m}_w \equiv \eta \dot{m}_{\ast}$, then we can write
\begin{equation}
	f_{\ast} = \frac{\dot{m}_{\ast}}{\dot{m}_b} = \frac{1}{1 + \eta} .
\end{equation}
One can show that, for energy-conserving winds, $\eta \propto m_h^{2/3} (1+z)^{-1}$, while for momentum-conserving winds, $\eta \propto m_h^{1/3} (1+z)^{-1/2}$ \cite{Dayal2014,Furlanetto2017}. Though simple, these models provide some physically-motivated guidance for extrapolating models to higher redshifts and/or fainter objects than are probed by current surveys. Current measurements can still accommodate either scenario, largely due to (i) the small time baseline overwhich measurements are available and (ii) uncertainties in correcting for dust reddening (see \S\ref{sec:dust}).

There are numerous other techniques that are commonly employed to model star formation in high-$z$ galaxies. For example, one need not require that star formation operate in an equilibrium with inflow and outflow, in which case Eq. \ref{eq:SFE_M} may be a more sensible choice than \ref{eq:SFE_MAR}. Many efforts are now underway to simulate galaxy formation using \textit{ab initio} cosmological simulations \cite{Vogelsberger2014,Schaye2015,Hopkins2014,OShea2015,Gnedin2014}, rather than using analytic or semi-analytic models. However, doing so self-consistently in statistically representative volumes is exceedingly computationally challenging, as a result, semi-analytic and semi-empirical prescriptions for star formation in reionization modeling remain the norm in 21-cm modeling codes \cite{Mirocha2017,Park2019,Mutch2016}. Though such approaches lack the spatial resolution to model individual galaxies or even groups of galaxies, including some information about the galaxy population permits joint modeling of 21-cm observables as well as high-$z$ galaxy luminosity functions, stellar mass functions, and so on, and thus open up the possibility of tightening constraints on the properties of galaxies using a multi-wavelength approach.

%%
% PopIII
%%
\subsubsection{Pop~III star formation} \label{sec:popIII}
The very first generations of stars to form in the Universe did so under very different conditions than stars today, so it is not clear that the star formation models outlined in the previous section apply. The first stars, by definition, formed from chemically-pristine material, since no previous generations of stars had existed to enrich the medium with heavy elements. This has long been recognized as a reason that the first stars are likely different than stars today \cite{Abel2000,Bromm1999,OShea2007,Yoshida2003}. Without the energetically low-lying electronic transitions common in heavy elements, hydrogen-only gas clouds cannot cool efficiently, as collisions energetic enough to excite atoms from $n=1$ to $n=2$ (which subsequently cool via spontaneous emission of $\Lya$ photons) imply temperatures of $\sim 10^4$ K, corresponding to virial masses of order $\sim 10^8 \ M_{\odot}$\footnote{Setting $T_{\min} \sim 10^4$ K (or $M_{\min} \sim 10^8 \ M_{\odot}$) is thus a way to roughly exclude the effects of PopIII-hosting ``minihalos'' in a 21-cm model.}. Such halos are increasingly rare at $z \gtrsim 10$.

Even in the absence of metals, there are cooling channels available in halos too small to support atomic (hydrogen) line cooling, i.e., with masses $M_{\odot} \lesssim 10^8 \ M_{\odot}$. Hydrogen molecules, $\Htwo$, can form using free electrons as a catalyst\footnote{Dust is the primary catalyst of $\Htwo$ formation in the local Universe, but of course is does not exist in the first collapsing clouds.}, 
\begin{align}
	\Hatom + e^- & \rightarrow \Hatom^- + h\nu \\
	\Hatom^- + \Hatom & \rightarrow \Htwo + e^- ,
\end{align}
These reactions are limited by the availability of free electrons\footnote{
Exotic models in which an X-ray background emerges before the formation of the first stars may affect early star formation by boosting the electron fraction.} and the survivability of $\Hminus$ ions. Even in the absence of astrophysical backgrounds, the formation of $\Htwo$ is limited by the CMB, which at the high redshifts of interest can dissociate the $\Hminus$ ion. \cite{Tegmark1997} found that the molecular hydrogen fraction in high-$z$ halos scales with the virial temperature as
\begin{equation}
	f_{\Htwo} \approx 3.5 \times 10^{-4} \left(\frac{\Tvir}{10^3 \ \mathrm{K}} \right)^{1.52} .
\end{equation}
Once the first stars form, the situation grows considerably more complicated. As will be detailed in the following section (\S\ref{sec:UV}), massive stars are prodigious sources of UV photons. Some of these photons originate in the Lyman-Werner band ($\sim 11.2$-$13.6$ eV), and are thus capable of dissociating molecular hydrogen. This processs is expected to quickly surpass $\Hminus$ dissociation by the CMB as the most important mechanism capable of regulating star formation in chemically pristine halos\footnote{If the PopIII IMF is very bottom-heavy, the resulting IR background  could continue to regulate star formation via $\Hminus$ photo-detachment \cite{WolcottGreen2012}.}. 

A substantial literature has emerged in the last $\sim 20$ years aimed at understanding the critical LW background intensity, $\JLW$, required to prevent star formation in high-$z$ minihaloes. For example, \cite{Visbal2014} find
\begin{equation}
	M_{\mathrm{crit}} = 2.5 \times 10^5 \left(\frac{1+z}{26} \right)^{-3/2} (1 + 6.96(4\pi \JLW)^{0.47})  \ \Msun
\end{equation}
where $\JLW$ is the LW background intensity in units of $10^{-21} \ \mathrm{erg} \ \mathrm{s}^{-1} \ \mathrm{cm}^{-2} \ \mathrm{Hz}^{-1} \ \mathrm{sr}^{-1}$. In principle, $M_{\mathrm{crit}}$ varies across the Universe from region to region as a function of the local LW intensity, but it is common to use the mean LW background intensity for simplicity. Note finally that sufficiently dense clouds can self-shield themselves against LW radiation, which is an important (and still uncertain) aspect of modeling LW feedback \cite{WolcottGreen2011}.

While the LW background is responsible for setting the \textit{minimum} halo mass required to host star formation, the \textit{maximum} mass of Pop~III halos, i.e., the mass at which halos transition from Pop~III to Pop~II star formation, depends on the interplay of many complex processes. For example, Pop~III supernovae will inject metals into the ISM of their host galaxies, which can trigger the transition to Pop~II star formation provided that at least some metals are retained and efficiently mix into proto-stellar clouds. The timescales involved are highly uncertain and may vary from halo to halo. Some halos may even be externally enriched \cite{Smith2015}. As a result, whereas UVLFs at high-$z$ provide some insight into the Pop~II SFE, the Pop~III SFE, which encodes the complex feedback processes at play, is completely unconstrained. 

\begin{figure*}[]
\begin{center}
\includegraphics[width=0.98\textwidth]{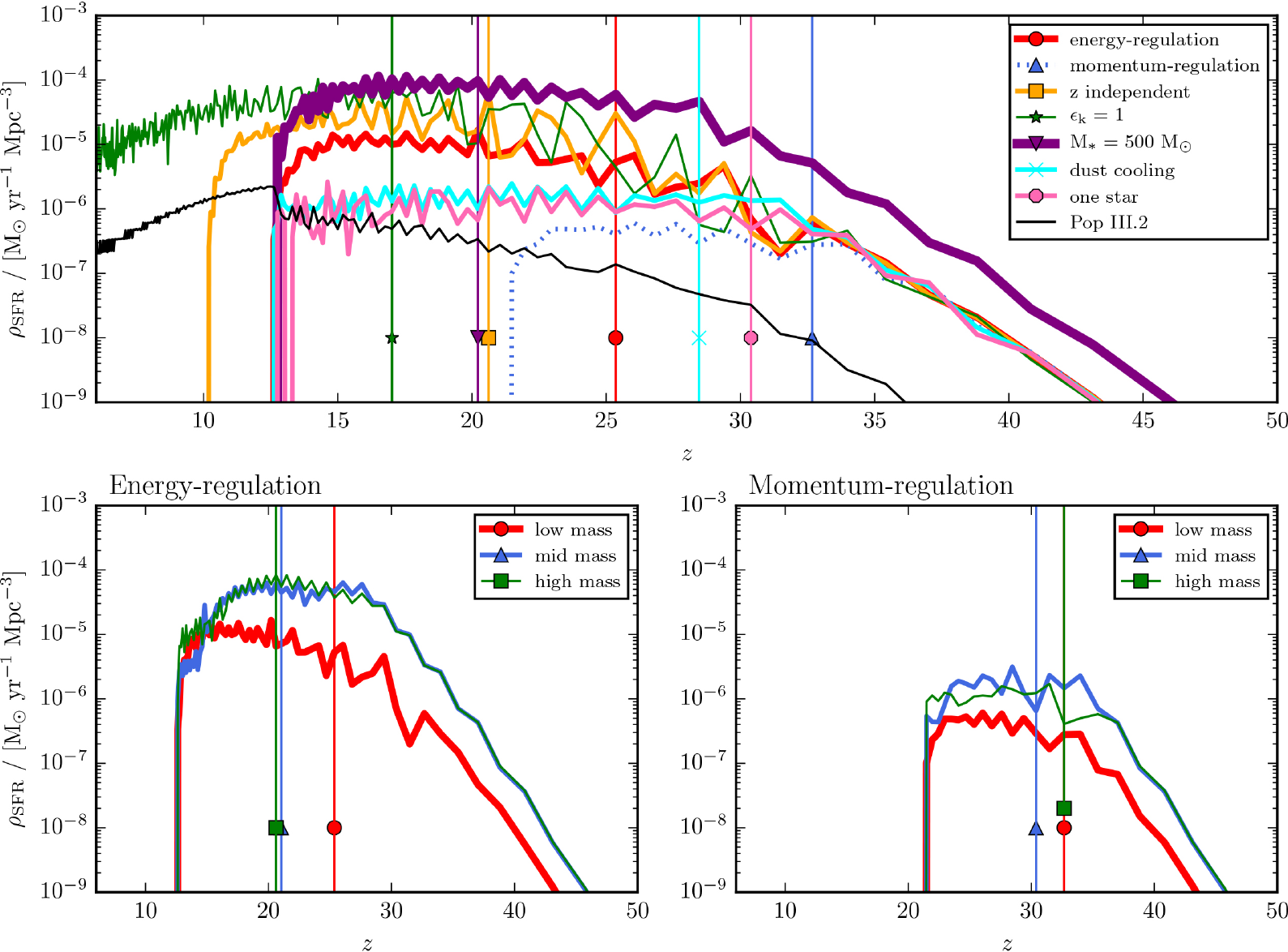}
\end{center}
\caption{{\bf Predictions for the Pop~III SFRD at high-$z$ \cite{Mebane2018}.} Upper panel shows results assuming a low-mass Salpeter-like Pop~III IMF with different assumptions about the Pop~II and Pop~III star formation prescriptions. Lower panels study the effects of Pop~III IMF between energy-regulated (left) and momentum-regulated (right) stellar feedback from Pop~II stars. Note that the Pop~II SFRD inferred from current UVLF measurements is of order $\sim 10^{-2} \ \sfrdunits$ at $z \sim 6$ \cite{Bouwens2015,Finkelstein2015}.}
\label{fig:popIII_sfrd}
\end{figure*}

Figure \ref{fig:popIII_sfrd} shows some example predictions for the Pop~III SFRD in a semi-analytic model of Pop~III star formation. Clearly, the level of Pop~III star formation is subject to many unknowns, which results in a vast array of predictions spanning spanning $\sim 3$ orders of magnitude in peak SFRD. This range is representative of the broader literature \cite{Trenti2009,Maio2010,Crosby2013,Visbal2018,Mebane2018,Jaacks2018,Sarmento2018}, with some studies favoring even slightly higher peak Pop~III SFRDs $\dot{\rho}_{\ast,\textsc{iii}} \sim 10^{-3} \ \sfrdunits$. Even a crude constraint on the Pop~III SFRD would rule out entire classes of models and thus provide a vital constraint on star formation processes in the earliest halos.

In 21-cm models it is common to neglect a detailed treatment of individual Pop~III star-forming halos, and instead parameterize the impact of Pop~II and Pop~III halos separately. One way to do this is to assume all atomic cooling halos form Pop~II stars (with $\zeta_{LW,\textsc{ii}}$, $\zeta_{X,\textsc{ii}}$, etc.), and all minihalos form Pop~III, with their own efficiency factors $\zeta_{X,\textsc{ii}}$ etc., and $\mmin$ determined self-consistently from the emergent LW background intensity \cite{Fialkov2014a,Mirocha2018}. We will revisit the predictions of these models in \S\ref{sec:predictions}.

% STARS
\subsection{UV Emission from Stars} \label{sec:UV}
With some handle on the efficiency of star formation, we now turn our attention to the efficiency with which stars generate UV photons, particularly in the Lyman continuum and Lyman Werner bands\footnote{We use this definition here loosely. Technically, the LW band is $\sim 11.2-13.6$ eV, a range which bounds photons capable of photo-dissociating molecular hydrogen, $H_2$. The $\Lya$ background is sourced by photons in a slightly broader interval, $\sim 10.2-13.6$ eV, but it is tedious to continually indicate this distinction, and as a result, we use ``LW band'' to mean all photons capable of eventually generating $\Lya$ photons.}. In principle the 21-cm background is also sensitive to the spectrum of even harder He-ionizing photons, since photo-electrons generated from helium ionization can heat and ionize the gas, while HeII recombinations can result in H-ionizing photons. The 21-cm signal could in principle also constrain the rest-frame infrared spectrum of stars in the early Universe, since IR photons can feedback on star-formation at very early times through $\Hminus$ photo-detachment \cite{WolcottGreen2012}. However, in this section, we focus only on the soft UV spectrum ($10.2 \lesssim E \lesssim 54.4$ eV) to which the 21-cm background is most sensitive.

The most detailed predictions for stellar spectra come from stellar population synthesis (SPS) models, which, as the name suggests, synthesize the spectra of entire stellar populations as a function of time. The key inputs to such models are:
\begin{itemize}
	\item The stellar initial mass function (IMF), $\xi(m)$, i.e., the number of relative number of stars formed in different mass bins. Commonly-adopted IMFs include Salpeter \cite{Salpeter1955}, Chabrier \cite{Chabrier2003}, Kroupa \cite{Kroupa2001}, and Scalo \cite{Scalo1998} which are all generally power-laws with indices $\sim -2.3$ , but differing in shape at the low mass end of the distribution ($M_{\ast} < 0.5 \ \Msun$ ).
	\item Models for stellar evolution, i.e., how stars of different masses traverse the Hertzprung-Russell diagram over time.
	\item Models for stellar atmospheres, i.e., predictions for the spectrum of individual stars as a function of their mass, age, and composition.
\end{itemize}
With all these ingredients, one can synthesize the spectrum of a stellar population formed some time $t$ after a ``burst,''
\begin{equation}
	L_{\nu}(t) = \int_0^t dt^{\prime} \int_{\mmin}^{\infty} dm \xi(m) l_{\nu} (m, t^{\prime}) \label{eq:Lcluster}
\end{equation}
where $l_{\nu}(m, t)$ is the specific luminosity of a star of mass $m$ and age $t$, and we have assumed that $\xi$ is normalized to the mass of the star cluster, $\int dm \xi(m) = M_{\ast}$. Equation \ref{eq:Lcluster} can be generalized to determine the spectrum of a galaxy with an arbitrary star formation history (SFH) composed of such bursts. Widely used stellar synthesis codes include \textsc{starburst99} \cite{Leitherer1999}, \textsc{bpass} \cite{Eldridge2009}, Flexible Stellar Population Synthesis (\textsc{fsps}) \cite{Conroy2009}, Stochastically Lighting up Galaxies (\textsc{slug}) \cite{daSilva2012}, and the Bruzual \& Charlot models \cite{Bruzual2003}.

Generally, 21-cm models do not operate at level of SPS models because the 21-cm background is insensitive to the detailed spectra and SFHs of individual galaxies. Instead, because 21-cm measurements probe the relatively narrow intervals $10.2 < h\nu / \mathrm{eV} < 13.6$ via Wouthuysen-Field coupling and $h\nu > 13.6$ eV through the ionization field, it is common to distill the predictions of SPS models into just two numbers, $\Nion$ and $N_{\alpha}$, which integrate over age and the details of the stellar SED, i.e.,
\begin{align}
	\Nion & = m_{\ast}^{-1} \int_0^{\infty} dt^{\prime} \int_{\nuLL}^{\infty} \frac{d\nu}{h\nu} L_{\nu}(t^{\prime}) \\
	N_{\alpha} & = m_{\ast}^{-1} \int_0^{\infty} dt^{\prime} \int_{\nuLya}^{\nuLL} \frac{d\nu}{h\nu} L_{\nu}(t^{\prime})
\end{align}
where $\nuLL$ is the frequency of the Lyman limit (13.6 eV) and $\nuLya$ is the Ly-$\alpha$ frequency. UV emission is dominated by massive, short-lived stars, hence the integration from $t=0$ to $t=\infty$. 

Assuming a Scalo IMF, stellar metallicity of $Z=Z_{\odot}/20$, using \textsc{starburst99} SPS model, \cite{Barkana2005} report $\Nion=4000$ and $N_{\alpha}=9690$ photons per baryon, the latter broken down further into sub-intervals between each Ly-$n$ resonance. The general expectation is for $\Nion$ and $N_{\alpha}$ increase for more top-heavy IMF and lower metallicity, meaning these values are likely to increase for Pop~III stars \cite{Bromm2001,Tumlinson2000,Schaerer2002}. Similarly, binary evolution can effectively increase the lifetimes of massive stars, leading to a net gain in UV photon production \cite{Stanway2016}.

Note that if simultaneously modeling the UVLF one generally assumes a constant star formation rate, in which case the UV luminosity of stellar populations asymptotes after a few hundred Myr. As a result, it is common to use the results of SPS models (with continuous star formation) at $t=100$ Myr when converting UV luminosity to SFR, as in Eq. \ref{eq:L1600}, though in reality the detailed star formation history is (generally) unknown. In this particular case, we do not quantify the UV luminosity in photons per stellar baryon -- instead, the exact values of $f_{\ast}$, $l_{1600}$, $\Nion$, and $N_{\alpha}$ should all be determined self-consistently from the model calibration. In other words, one cannot simply change $\Nion$ and $f_{\ast}$ independently, since the inferred value of $f_{\ast}$ depends on the assumed stellar population model, and thus implicitly on $\Nion$ (see also \S\ref{sec:predictions} and Fig.  \ref{fig:gs_metallicity}; \cite{Mirocha2017}).

%%
% DUST
%%
\subsection{Attenuation of Stellar UV Emission by Dust} \label{sec:dust}
Even if the intrinsic stellar spectrum of galaxies were known perfectly, our ability to draw inferences about star formation are hampered by the presence of dust, which dims and reddens the ``true'' spectrum of galaxies. The opacity of dust is an inverse function of wavelength, meaning its impact is greatest at short wavelengths \cite{Weingartner2001}. Unfortunately, most observations of high-$z$ galaxies (so far) target the rest UV spectrum of galaxies\footnote{This is simply due to the limited availability and sensitivity of near-infrared observations, which will soon be greatly enhanced by the James Webb Space Telescope.}, and thus must be considerably ``dust corrected'' before star formation rates and/or efficiencies are estimated. 

However, correcting for the effets of dust attenuation is not completely hopeless. If we assume that UV-heated dust grains radiate as blackbodies, we would expect to see an ``infrared excess'' (IRX) in galaxies with redder-than-expected UV continuua, as UV reddening is suggestive of dust attenuation. If we assume for simplicity a power-law UV continuum, $f_{\lambda} = f_{\lambda,0} \lambda^{\beta}$, we would expect an excess
\begin{equation}
	\mathrm{IRX}_{1600} \equiv \frac{F_{\mathrm{FIR}}}{F_{1600}} = \frac{\int_{912}^{\infty} f_{\lambda,0} (1 - e^{-\tau_{\lambda}}) d\lambda}{f_{1600,0} e^{-\tau_{\lambda}}} \left(\frac{F_{\mathrm{FIR}}}{F_{\mathrm{bol}}} \right)
\end{equation}
where $\tau_{\lambda}$ is the wavelength-dependent dust opacity, equivalently written via $10^{-0.4 A_{1600}} = e^{-\tau_{1600}}$, where $A_{1600}$ is the extinction at 1600$\AA$ in magnitudes, $f_{1600,0}$ is the intrinsic intensity at $1600 \AA$, and $F_{\mathrm{FIR}}/F_{\mathrm{bol}}$ is a correction factor that accounts for the fraction of the bolometric dust luminosity emitted in the far-infrared band of observation\footnote{Note that the above expression assumes that all heating is done by photons redward of the Lyman limit and neglects heating by line photons.}. 

An empirical constraint on the so-called IRX-$\beta$ relation was first presented by \cite{Meurer1999}, who found $A_{1600} = 4.43 - 1.99 \beta_{\mathrm{obs}}$, where $\beta_{\mathrm{obs}}$ is the logarithmic slope of the observed rest UV spectrum. The intrinsic UV slope of young stellar populations is generally $-3 \lesssim \beta \lesssim -2$, which, coupled with the fact that observed slopes can be $-2 \lesssim \beta_{\mathrm{obs}} \lesssim -1$ \cite{Finkelstein2012,Bouwens2014}, implies that corrections of potentially several magnitudes are likely in order. There is an ongoing debate in the field regarding the origin of the IRX-$\beta$ relation, the cause of its scatter, and the possibility that it evolves with time \cite{Narayanan2018,Salim2019}. Early efforts with ALMA \cite{Capak2015,Bouwens2016} are beginning to test these ideas with rest-IR observations of $z \sim 6$ star-forming galaxies, with some hints that there is evolution in IRX-$\beta$ indicative of reduced dust obscuration in higher redshift galaxies. However, such inferences are currently dependent on assumptions for the dust temperature -- if dust at high-$z$ is warmer than dust at low-$z$, the data may be consistent with no evolution in $A_{1600}$ at fixed $M_{\mathrm{UV}}$ or stellar mass. These uncertainties in correcting high-$z$ rest-UV measurements for dust reddening may affect the normalization of $f_{\ast}$ at the factor of $\sim$ few level, and could also bias the shape of the inferred $f_{\ast}$ depending on precisely how $A_{1600}$ scales with $\MUV$ (or $m_{\ast}$).

%%
% fesc
%%
\subsection{Escape of UV Photons from Galaxies} \label{sec:fesc}
While photons with wavelengths longer than $912 \ \AA$ are most likely to be absorbed by dust grains, as described in the previous section, photons with wavelengths shortward of $912 \ \AA$ will be absorbed by hydrogen and helium atoms. As a result, reionization models must also account for local attenuation, since the ionization state of intergalactic gas is of course only influenced only by the ionizing photons that are able to escape galaxies. The fraction of photons that escape relative to the total number produced is quantified by the escape fraction, $\fesc$, and is the final component of the ionizing efficiency, $\zeta$, introduced previously (see Eq. \ref{eq:zeta}).

Current constraints on high-$z$ galaxies and reionization suggest that $\fesc$ must be $\sim 10-20\%$ \cite{Robertson2015}. The result is model-dependent, however, as it relies on assumptions about the UV photon production efficiency in galaxies and extrapolations to source populations beyond current detection limits. For example, if $\fesc$ depends inversely on halo mass, reionization can be driven by galaxies that have yet to be detected directly \cite{Finkelstein2019}\footnote{This scenario is appealing because it can explain the very gradual evolution in the post-reionization ionizing background, and rarity of galaxies leaking LyC radiation at $3 \lesssim z \lesssim 6$ \cite{Shapley2006}}.

Numerical simulations now lend credence to the idea that escape fractions of $\sim 10-20$\% are possible, with perhaps even larger $\fesc$ in low-mass halos \cite{Kimm2014,Xu2016}. The basic trend is sensible: as the depth of halo potentials declines, supernovae explosions can more easily excavate clear channels through which photons escape. However, there is far from a consensus on this issue. For example, the \textsc{FIRE} simulations do not see evidence that $\fesc$ depends on halo mass \cite{Ma2015}, and on average $\fesc \lesssim 5$\%, causing some tension with reionization constraints unless binary models \cite{Eldridge2009} are employed. This result, coupled with the very high resolution in \textsc{FIRE}, is more suggestive of a scenario in which $\fesc$ is set by very small-scale structure in the ISM, rather than the depth of the host halo potential.

21-cm measurements in principle open a new window into constraining $\fesc$. If, for example, the UV/SFR conversion factor is well known and dust can be dealt with (see \S\ref{sec:UV}-\ref{sec:dust}), joint fitting 21-cm power spectra and high-$z$ galaxy LFs can isolate $f_{\ast}$ and $\fesc$ \cite{Park2019,Greig2019}. Note, however, that this is still model-dependent, as $f_{\ast}$ must be extrapolated to some limiting UV magnitude or halo mass in order to obtain the total photon production rate per unit volume. 

%%%
%% BHs ETC
%%%
\subsection{X-rays from Stellar-Mass Black Holes} \label{sec:hmxbs}
Though stars themselves emit few photons at energies above the HeII-ionizing edge ($\sim 54.4$ eV), their remnants can be strong X-ray sources and thus affect the IGM temperature. While solitary remnants will be unlikely to accrete much gas from the diffuse ISM (though see \S\ref{sec:smbhs}), remants in binary systems may accrete gas from their companions, either via Roche-lobe overflow or stellar winds. Such systems are known as X-ray binaries (XRBs), further categorized by the mass of the donor star: ``low-mass X-ray binaries'' (LMXBs) are those fueled by Roche-lobe overflow from a low-mass companion, while ``high-mass X-ray binaries'' (HMXBs) are fed by the winds of massive companions. XRBs exhibit a rich phenomenology of time- and frequency-dependent behavior and are thus interesting in their own right. For a review see, e.g., \cite{Remillard2006}.

In nearby star-forming galaxies, the X-ray luminosity is generally dominated by HMXBs \cite{Gilfanov2004,Fabbiano2006,Mineo2012a}. Furthermore, the total luminosity in HMXBs scales with the star formation rate, as expected given that the donor stars in these systems are massive, short-lived stars. An oft-used result in the 21-cm literature stems from the work of \cite{Mineo2012a}, who find
\begin{equation}
	L_X = 2.6 \times 10^{39} \left(\frac{\SFR}{M_{\odot} \ \mathrm{yr}^{-1}} \right) \ \mathrm{erg} \ \mathrm{s}^{-1} \label{eq:LxSFR_Mineo}
\end{equation}
where $L_X$ is defined here as the luminosity in the 0.5-8 keV band. This relation provides an initial guess for many 21-cm models, which add an extra factor $f_X$ to parameterize our ignorance of how this relation evolves with cosmic time. For example, \cite{Furlanetto2006} write
\begin{equation}
	L_X = 3 \times 10^{40} f_X \left(\frac{\SFR}{M_{\odot} \ \mathrm{yr}^{-1}} \right) \ \mathrm{erg} \ \mathrm{s}^{-1} , \label{eq:LxSFR_Furlanetto}
\end{equation}
which is simply Equation \ref{eq:LxSFR_Mineo} re-normalized to a broader energy range, $0.2 < h\nu/\mathrm{keV} < 3\times 10^4$, assuming a power-law spectrum with spectral index $\alpha_X=-1.5$, where $\alpha_X$ is defined by $L_E \propto E^{\alpha_X}$. Coupled with estimates for the star formation rate density at high-$z$, the $L_X$-SFR relation suggests that X-ray binaries could be considerable sources of heating in the high-$z$ IGM \cite{Furlanetto2006,Fragos2013,Mirocha2014,Fialkov2014b,Madau2017}.

The normalization of these empirical $L_X$-SFR relations are not entirely unexpected, at least at the order-of-magnitude level. For example, if one considers a galaxy forming stars at a constant rate, a fraction $f_{\bullet} \simeq 10^{-3}$ of stars will be massive enough ($M_{\ast} > 20 \ M_{\odot}$) to form a black hole assuming a Chabrier IMF. Of those, a fraction $\fbin$ will have binary companions, with a fraction $\fsurv$ surviving the explosion of the first star for a time $\tau$. If accretion onto these black holes occurs in an optically thin, geometrically-thin disk with radiative efficiency $\epsilon_{\bullet} = 0.1$ which obeys the Eddington limit, then a multi-color disk spectrum is appropriate and a fraction $f_{0.5-8}=0.84$ of the bolometric luminosity will originate in the 0.5-8 keV band. Finally, assuming these BHs are ``active'' for a fraction $\fact$ of the time, we can write \cite{Mirabel2011,Mirocha2018}
\begin{equation}
	L_X \sim 2 \times 10^{39} \mathrm{erg} \ \mathrm{s}^{-1} \left(\frac{\SFR}{\SFRunits} \right) \left(\frac{\epsilon_{\bullet}}{0.1}\right) \left(\frac{f_{\bullet}}{10^{-3}} \right) \left(\frac{\fbin}{0.5} \right) \left(\frac{\fsurv}{0.2} \right) \left(\frac{\tau}{20 \ \mathrm{Myr}} \right) \left(\frac{\fact}{0.1} \right) \left(\frac{f_{0.5-8}}{0.84} \right) . \label{eq:LxSFR}
\end{equation}
While several of these factors are uncertain, particularly $\fsurv$ and $\fact$, this expression provides useful guidance in setting expectations for high redshift. For example, it has long been predicted that the first generations of stars were more massive on average than stars today owing to inefficient cooling in their birth clouds. This would boost $f_{\bullet}$, and thus $L_X/\SFR$, so long as most stars are not in the pair-instability supernova (PISN) mass range, in which no remnants are expected. 

\begin{figure*}[]
\begin{center}
\includegraphics[width=0.49\textwidth]{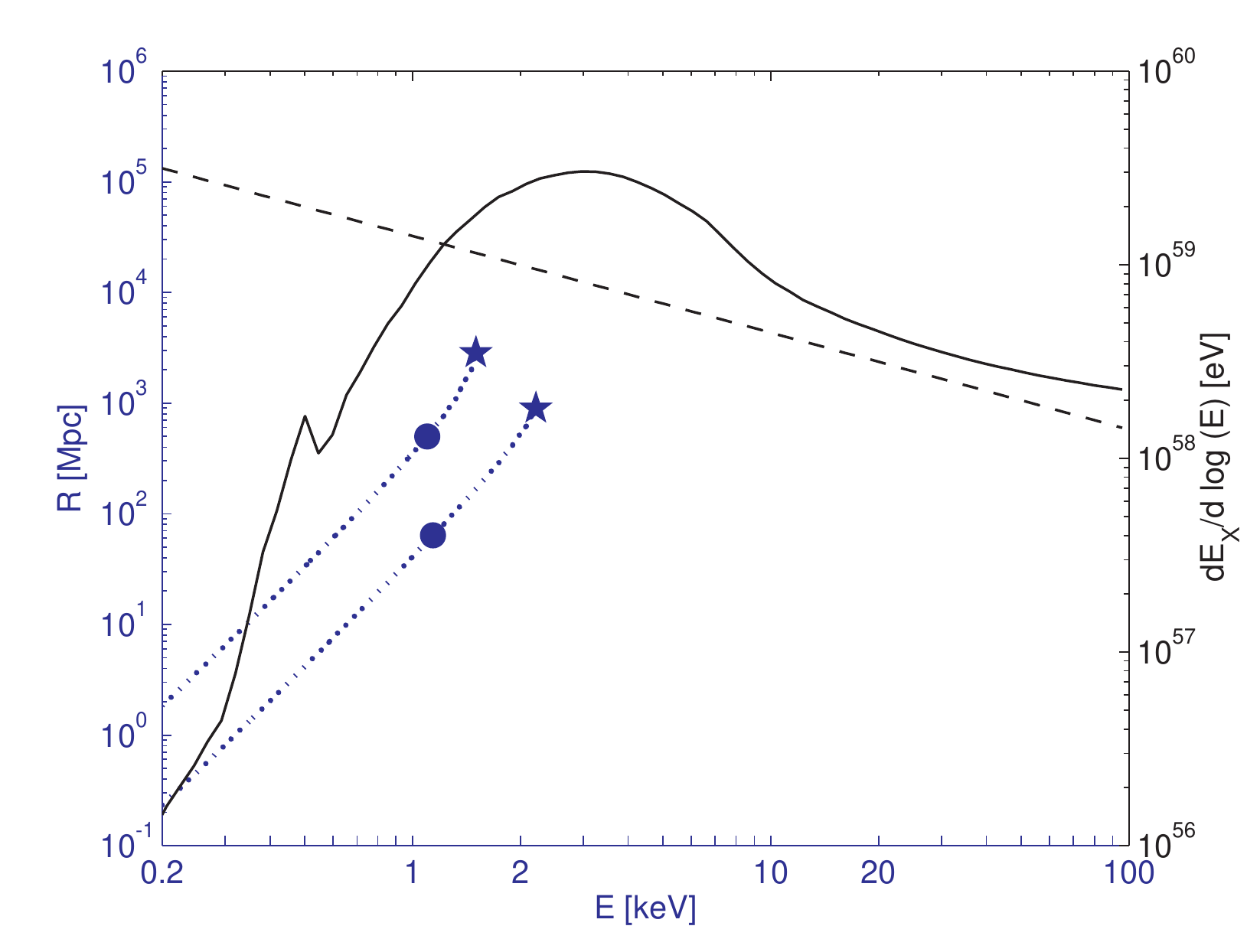}
\includegraphics[width=0.49\textwidth]{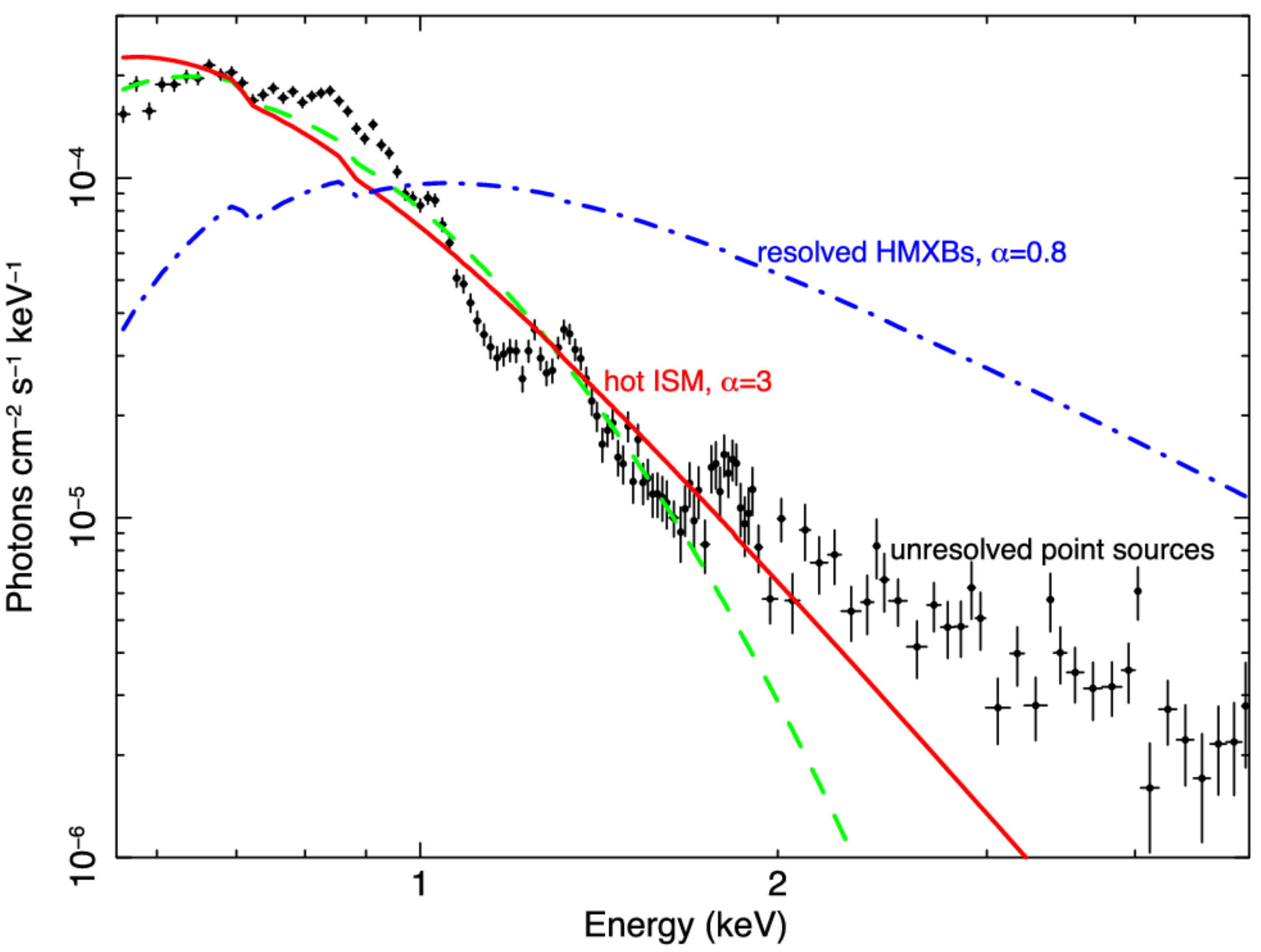}
\end{center}
\caption{{\bf Models for the X-ray spectra of star-forming galaxies.} \textit{Left:} Template Cygnus X-1 HMXB spectrum (solid) compared with power-law (dashed), with mean free path shown on left scale and relative luminosity on right \cite{Fialkov2014b}. \textit{Right:} Typical HMXB spectrum (blue) compared to soft X-ray spectrum characteristic of bremmstrahlung emission from hot ISM gas \cite{Pacucci2014}.}
\label{fig:xray_seds}
\end{figure*}

There are of course additional arguments not present in Eq. \ref{eq:LxSFR}. For example, the MCD spectrum is only a good representation of HMXB spectra in the ``high soft'' state. At other times, in the so-called ``low hard'' state, HMXB spectra are well fit by a power-law. The relative amount of time spent in each of these states is unknown. Figure \ref{fig:xray_seds} compares typical HMXB spectra with the spectrum expected from hot ISM gas (see \S\ref{sec:ism_bremms}).

In addition, physical models for the $L_X$-SFR relation may invoke the metallicity as a driver of changes in the relation with time and/or galaxy (stellar) mass. As the metallicity declines, one might expect the stellar IMF to change (as outlined above), however, the winds of massive stars responsible for transferring material to BHs will also grow weaker as the opacity of their atmospheres declines. As a result, increases in $L_X$/SFR likely saturate below some critical metallicity. Observations of nearby, metal-poor dwarf galaxies support this picture, with $L_X$/SFR reaching a maximum of $\sim 10$x the canonical relation quoted in Eq. \ref{eq:LxSFR_Mineo} \cite{Mineo2012a}.

%{\color{red} Left to discuss:
%\begin{itemize}
%	\item Observational limits on $L_X$/SFR from Chandra stacks?
%	\item Low metallicity constraints.
%\end{itemize}}

%%
% PopIII
%%
\subsubsection{X-rays from Super-Massive Black Holes} \label{sec:smbhs}
Though super-massive black holes (SMBHs) are exceedingly rare and thus unlikely to contribute substantially to the ionizing photon budget for reionization \cite{Hassan2018} (though see \cite{Madau2015}), fainter -- but more numerous -- intermediate mass black holes (IMBHs) with $10^3 \lesssim M_{\bullet} / M_{\odot} \lesssim 10^6$ could have a measureable impact on the IGM thermal history \cite{Zaroubi2007,Ripamonti2008,Tanaka2016}. Growing black holes, if similar to their low-$z$ counterparts, could also generate a strong enough radio background to amplify 21-cm signals \cite{EwallWice2018}, possibly providing an explanation for the anamalous depth of the EDGES global 21-cm measurement \cite{Bowman2018}. Such scenarios cannot yet be ruled out via independent measurements. For example, the unresolved fraction of the cosmic X-ray background still permits a substantial amount of accretion at $z \gtrsim 6$ \cite{McQuinn2012,Fialkov2017,Mirocha2018}, while just $\sim 10$\% of the radio excess reported by ARCADE-2 \cite{Fixsen2011,Singal2018} must originate at $z \gtrsim 18$ in order to explain the EDGES signal \cite{Feng2018}. Given the persistent challenge in explaining the existence of SMBHs at $z \gtrsim 6$, the signatures of BH growth in the 21-cm background are worth exploring in more detail.

% Hot gas
\subsection{X-rays from Shocks and Hot Gas} \label{sec:ism_bremms}
While compact remnants of massive stars are likely the leading producer of X-rays in high-$z$ star-forming galaxies, the supernovae events in which these objects are formed may not be far behind. Supernovae inject a tremendous amount of energy into the surrounding medium, which then cools either via inverse Compton emission (in supernova remnants; \cite{Oh2001}) or eventually via bremsstrahling radiation (in the hot interstellar medium; ISM). Because these sources are related to the deaths of massive stars their luminosity is expected to scale with SFR, as in the case of HMXBs. Indeed, \cite{Mineo2012b} find that diffuse X-ray emission in nearby sources follows the following relation in the 0.5-2 keV band:
\begin{equation}
	L_X = 8.3 \times 10^{38} \left(\frac{\SFR}{M_{\odot} \ \mathrm{yr}^{-1}} \right) \ \mathrm{erg} \ \mathrm{s}^{-1} \label{eq:LxSFRII_Mineo}
\end{equation}
This luminosity is that from all unresolved emission, and as a result, is not expected to trace emission from the hot ISM alone. Emission from supernova remnants will also contribute to this luminosity, as will fainter, unresolved HMXBs and LMXBs. \cite{Mineo2012b} estimate that $\sim 30-40$\% of this emission may be due to unresolved point sources.

Though the soft X-ray luminosity from hot gas appears to be subdominant to the HMXB component in nearby galaxies, at least in total power, there are of course uncertainties in how these relations evolve. Furthermore, the bremmstrahlung emission characteristic of hot ISM gas has a much steeper $\sim \nu^{-2.5}$ spectrum than inverse Compton ($\sim \nu^{-1}$) or XRBs ($\sim \nu^{-1}$ or $\nu^{-1.5}$), and thus may heat more efficiently (owing to $\sigma \propto \nu^{-3}$ cross section) provided soft X-rays can escape galaxies. 

%%
% NHI, Emin
%%
\subsection{Escape of X-rays from Galaxies} \label{sec:xray_esc}
Though the mean free paths of X-rays are longer than those of UV photons, they still may not all escape from galaxies into the IGM. For example, hydrodynamical simulations suggest typical hydrogen column densities of $\NHI \sim 10^{21} \ \mathrm{cm}^{-2}$ in low-mass halos \cite{Das2017}, which is substantial enough to eliminate emission below $\sim 0.5$ keV. 

Given the many unknowns regarding X-ray emission in the early Universe, 21-cm models often employ a three-parameter approach, i.e., instead of a single value of $\zeta_X$, the specific X-ray luminosity is modeled as 
\begin{equation}
	L_{X,\nu} = L_{X,0} \left(\frac{h \nu}{1 \mathrm{keV}} \right)^{\alpha_X} \exp\left[-\sigma_{\nu} \NHI \right]
\end{equation}
and the normalization, $L_{X,0}$, spectral index $\alpha_X$, and typical column density, $\NHI$, are left as free parameters. It is common to approximate this intrinsic attenuation with a piecewise model for $L_X$, i.e., 
\begin{align}
L_{X,\nu} = \left\{ \begin{array}{cc} 
                0 & h\nu < \Emin \\
                L_{X,0} \left(\frac{h \nu}{1 \mathrm{keV}} \right)^{\alpha_X} & h\nu \geq \Emin \
                \end{array} \right.
\end{align}
Note that $\NHI$ (or $\Emin$) can be degenerate with the intrinsic spectrum, e.g., the SED of HMXBs in the high-soft state exhibits a turn-over at energies $h\nu < 1$ keV, which could be mistaken for strong intrinsic absorption \cite{Mirocha2014}.

% DCBHS? 
\subsection{Cosmic Rays from Supernovae}
High energy cosmic rays (CRs) produced in supernovae explosions offer another potential source of ionization and heating in the bulk IGM \cite{Nath1993,Sazonov2015,Leite2017}), though most likely the effects are only discernible in the thermal history. Simple models suggest that CRs can raise the IGM temperature by $\sim 10-200$ K by $z \sim 10$ depending on the details of the CR spectrum \cite{Leite2017}. CRs are thus a potentially important, though relatively unexplored, source of heating in the high-$z$ IGM.

%%%
%% Modeling 
%%%
\section{Predictions for the 21-cm Background} \label{sec:predictions}
So far we have assembled a simple physical picture of the IGM at high redshift (\S\ref{sec:igm}) and the sources most likely to affect its properties (\S\ref{sec:sources}). Here, we finally describe the generic sequence of events predicted in most 21-cm models and the sensitivity of the 21-cm background to various model parameters of interest. 

\begin{figure*}[]
\begin{center}
\includegraphics[width=0.98\textwidth]{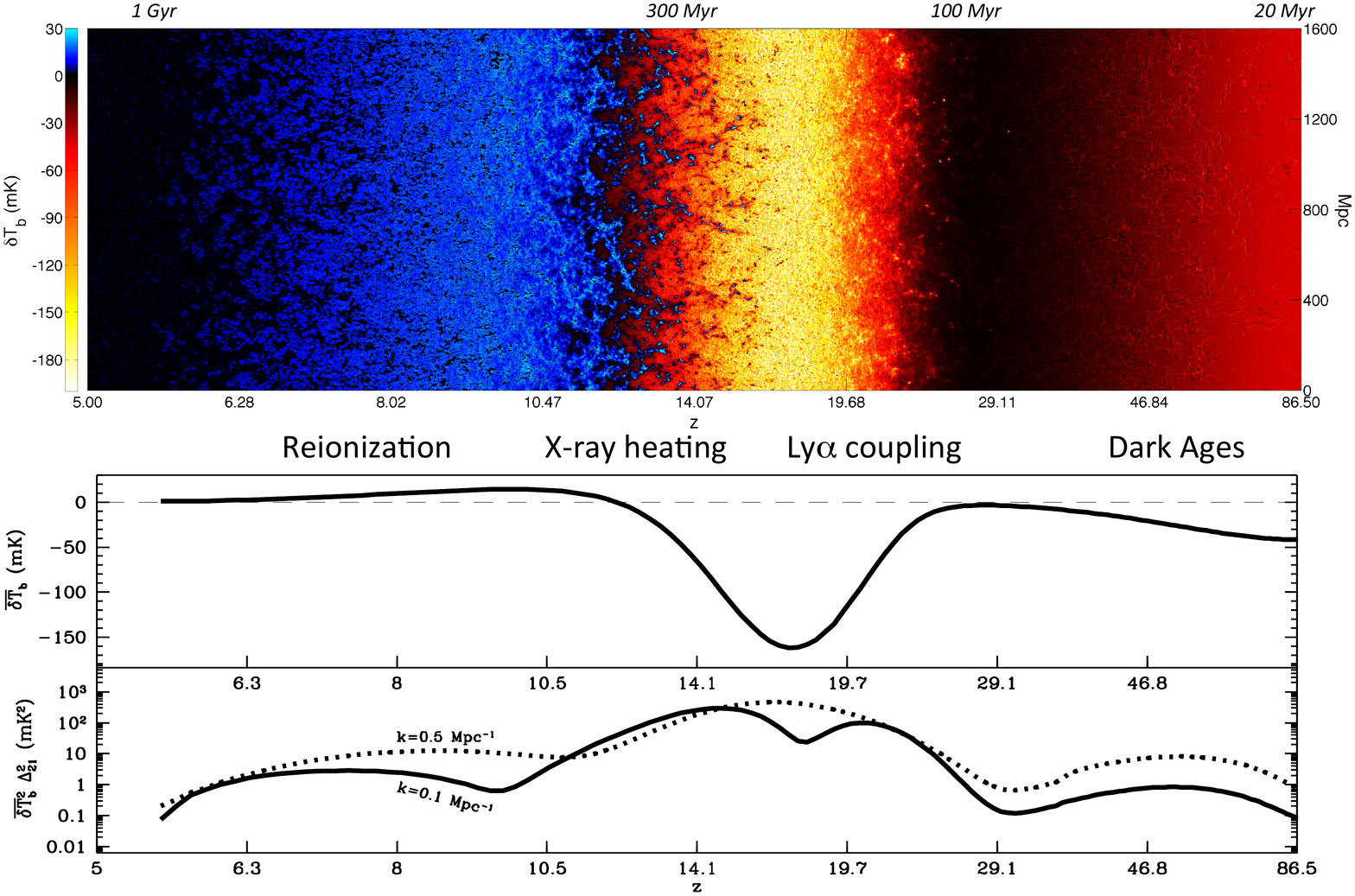}
\end{center}
\caption{{\bf Predictions from the \textsc{21cmfast} Evolution of Structure (EoS) model suite \cite{Mesinger2016}.} \textit{Top}: 2-D slice of the brightness temperature field, with red colors indicating a cool IGM, blue colors indicative of a heated IGM, and black representing a null signal (either due to ionization or $\TS=\Tcmb$). \textit{Middle:} Global 21-cm signal, with dashed line indicating  $\delta T_b = 0$. \textit{Bottom:} Evolution of the dimensionless 21-cm power spectrum, $\Delta^2 = k^3 P(k) / 2\pi^2$, on two different scales, $k=0.5 \ \mathrm{Mpc}^{-1}$ (dotted) and $k=0.1 \ \mathrm{Mpc}^{-1}$ (solid).}
\label{fig:eosFG}
\end{figure*}

Figure \ref{fig:eosFG} shows an illustrative example using \textsc{21cmfast} \cite{Mesinger2011} including a 2-D slice of the $\delta T_b$ field, the global 21-cm signal, and power spectrum on two spatial scales \cite{Mesinger2016}. Time proceeds from right to left from $\sim 20$ Myr after the Big Bang until the end of reionization $\sim 1$ Gyr later.

There are four distinct epochs indicated within this time period, which we describe in more detail below.
\begin{description}
	\item[The Dark Ages:] As the Universe expands after cosmological recombination, Compton scattering between free electrons and photons keeps the radiation and matter temperature in equilibrium. The density is high enough that collisional coupling remains effective, and so $\TS = \TK = \TCMB$. Eventually, Compton scattering becomes inefficient as the CMB cools and the density continues to fall, which allows the gas to cool faster than the CMB. Collisional coupling remains effective for a short time longer and so $\TK$ initially tracks $\TS$. This results in the first decoupling of $\TS$ from $\TCMB$ at $z \sim 150$, resulting in an absorption signature at $z \sim 80$ ($\nu \sim 15$ MHz), which comes to an end as collisional coupling becomes inefficient, leaving $\TS$ to reflect $\TCMB$ once again.
	\item[Ly-$\alpha$ coupling:] When the first stars form they flood the IGM with UV photons for the first time. While Lyman continuum photons are trapped near sources, photons with energies $10.2 < h\nu / \mathrm{eV} < 13.6$ either redshift directly through the $\Lya$ resonance or cascade via higher $\Lyn$ levels, giving rise to a large-scale $\Lya$ background capable of triggeiring Wouthuysen-Field coupling as they scatter through the medium (see also Ch. 1 and \S\ref{sec:lya}). As a result, $\TS$ is driven back toward $\TK$, which (in most models) still reflects the cold temperatures of an adiabatically-cooling IGM.
	\item[X-ray Heating:] The first generations of stars beget the first generations of X-ray sources, whether they be the explosions of the first stars themselves or remnant neutron stars or black holes that subsequently accrete. Though the details change depending on the identity of the first X-ray sources (see \S\ref{sec:hmxbs}-\ref{sec:ism_bremms}), generally such sources provide photons energetic enough to travel great distances. Upon absorption, they heat and partially ionize the gas, eventually driving $\TS > \TCMB$. Once $\TS \gg \TCMB$, the 21-cm signal ``saturates,'' and subsequently is sensitive only to the density and ionization fields. However, it is possible that heating is never ``complete'' in this sense before reionization, meaning neutral pockets of IGM gas may remain at temperatures at or below $\TCMB$ until they are finally engulfed by ionized bubbles.
	\item[Reionization:] As the global star formation rate density climbs, the growth of ionized regions around groups and clusters of galaxies will continue, eventually culminating in the completion of cosmic reionization. This rise in ionization corresponds to a decline in the amount of neutral hydrogen in the Universe capable of producing generating 21-cm signals. As a result, the amplitude of the 21-cm signal, both in its mean and fluctuations, falls as reionization progresses. After reionization, neutral hydrogen only remains in systems over-dense enough to self-shield from the UV background.
\end{description}

The particular model shown in Figure \ref{fig:eosFG} \cite{Mesinger2016} assumes that very faint galaxies dominate the UV and X-ray emissivity, which results in relatively early features in the 21-cm background, e.g., both the power spectrum and global 21-cm signal peak in amplitude at $z \sim 18$. Reionization and reheating occur later in scenarios in which more massive halos dominate the emissivity, and may even occur simultaneously, resulting in strong 21-cm signals at $z \lesssim 12$ \cite{Mesinger2016,Mirocha2017,Park2019}. 

For the remainder of this section we focus on changes in the 21-cm signal wrought by parameters of interest. We limit our discussion to the global 21-cm signal and power spectrum, though there are of course many other statistics one could use to constrain model parameters (see Chapter 4). We note that there is no consensus parameterization for models of galaxy formation or the 21-cm background, nor do all models incorporate the same physical processes or employ the same numerical techniques. As a result, in this section we make no effort to closely compare or homogenize results from the literature, but instead draw examples from many works in order to illustrate different aspects of the 21-cm background as a probe of galaxy formation. 

% zeta
\subsection{Dependence on the Ionizing Efficiency} \label{sec:dep_zeta}
Generally written as $\zeta$ or $\zetaI$, the ionizing efficiency quantifes the number of Lyman continuum (LyC) photons that are produced in galaxies and escape into the IGM, i.e., $\zeta=f_{\ast} \Nion f_{\esc}$ (see \S\ref{sec:sfe}-\ref{sec:fesc}). As a result, this parameter affects primarily the lowest redshifts $6 \lesssim z \lesssim 10$ ($\nu \gtrsim 130$ MHz), during which the bulk of reionization likely takes place.

\begin{figure}[]
\begin{center}
\includegraphics[width=0.3\textwidth]{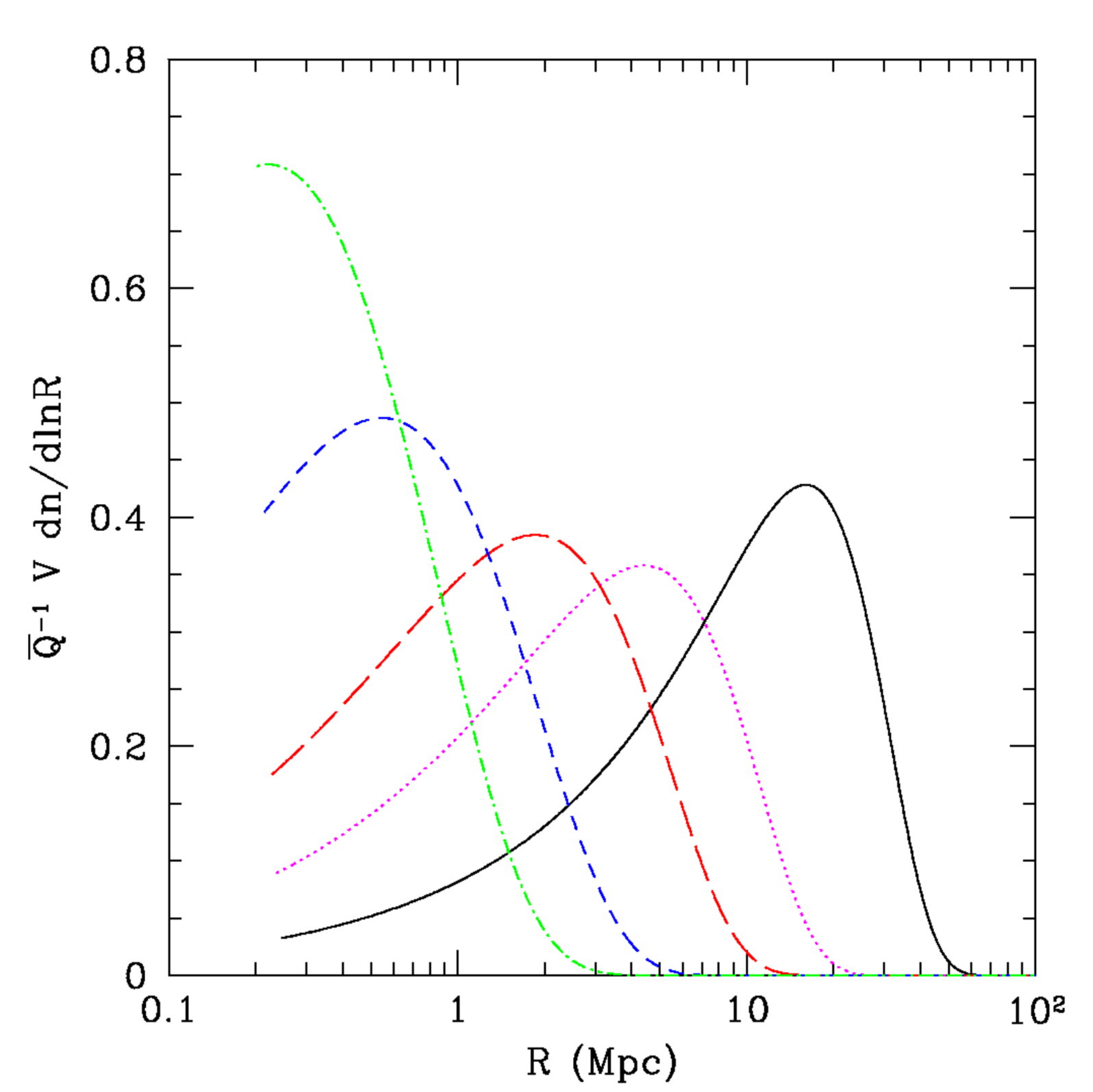} \includegraphics[width=0.3\textwidth]{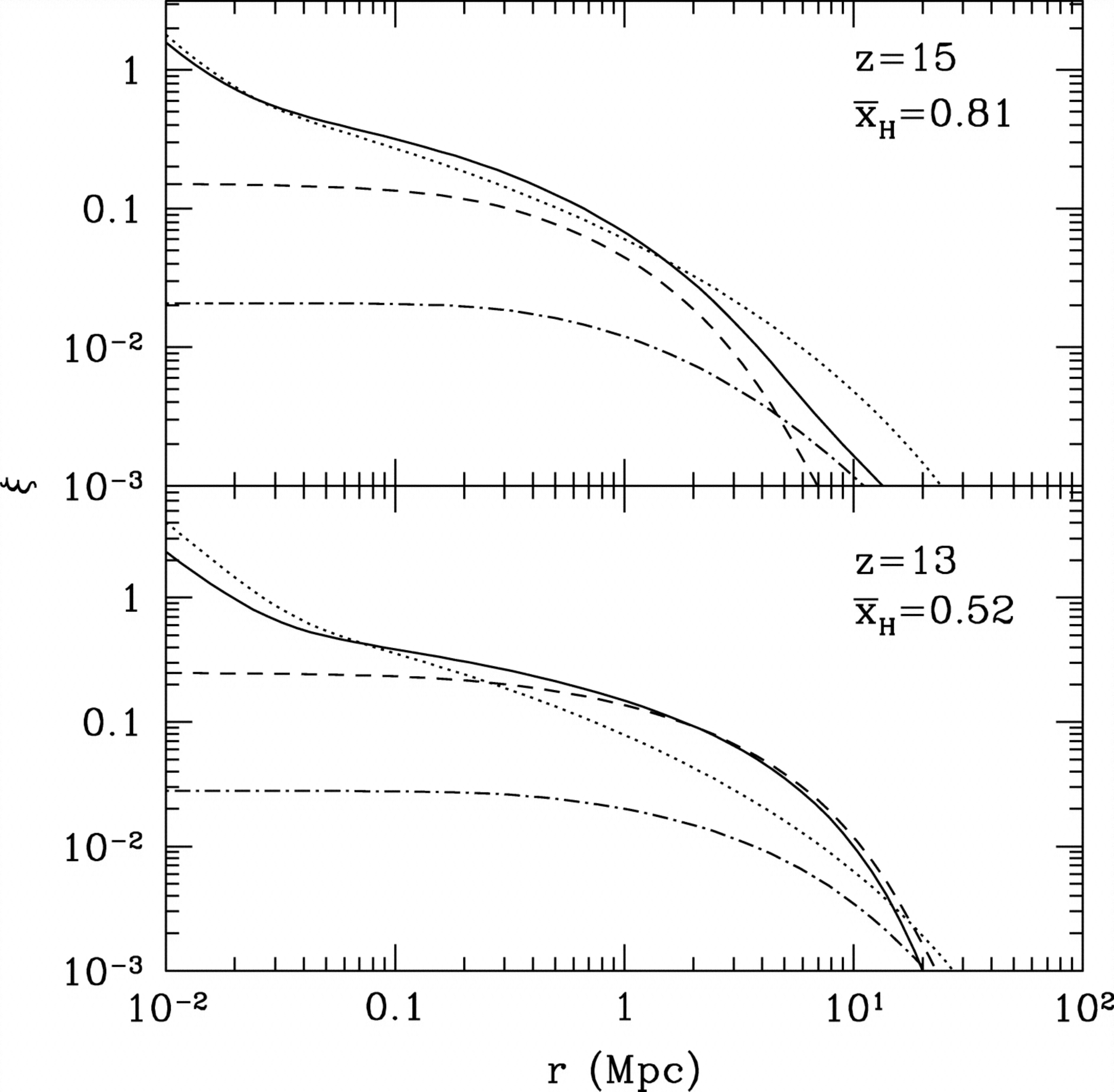} \includegraphics[width=0.3\textwidth]{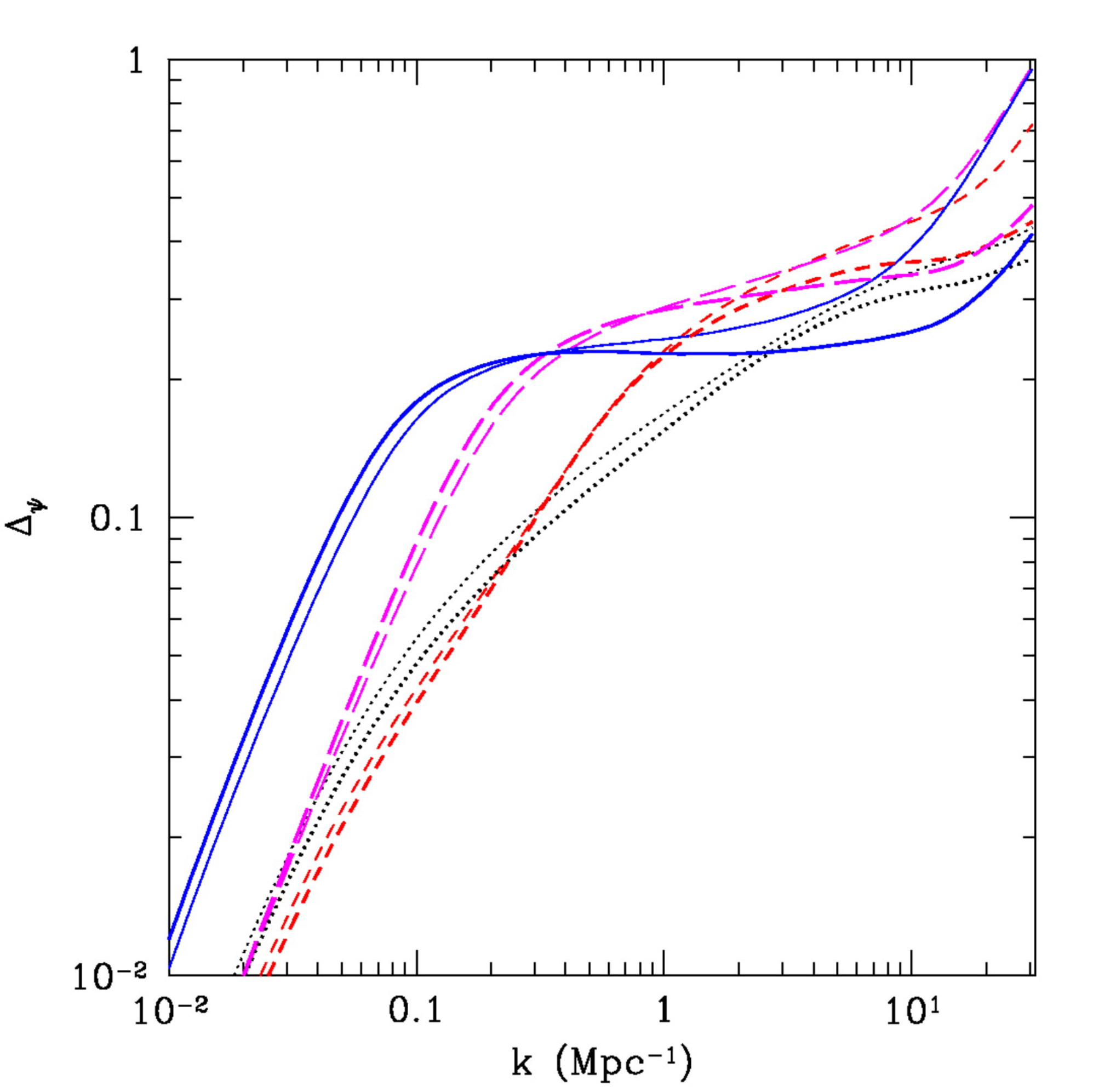}
\end{center}
\caption{{\bf Analytic models of bubble growth during the EoR \cite{Furlanetto2004}.} \textit{Left:} Bubble size distributions at bubble filling fractions of $Q=0.037, 0.11, 0.3, 0.5$, and 0.74, from left to right. \textit{Middle:} Correlation function of $\psi = \xHI (1 + \delta)$ (solid) at $\xHI=0.81$ (top) and $\xHI = 0.52$ (bottom) as well as its constituent components, including the ionization auto-correlation function (dashed), density auto-correlation function (dotted), and cross-correlation function between ionization and density (dot-dashed). \textit{Right:} Dimensionless power spectrum of $\psi$ for different values of $\zeta$, including $\zeta=12$ (thin) and $\zeta=40$ (thick), at several neutral fractions, $\xHI=0.96$ (dotted), 0.8 (short-dashed), 0.5 (long-dashed), and 0.26 (solid).}
\label{fig:fzh04}
\end{figure}

Figure \ref{fig:fzh04} shows predictions for the growth of ionized bubbles in the excursion set formalism \cite{Furlanetto2004}. In time, bubbles grow larger, eventually reaching typical sizes of $\sim $tens of Mpc during reionization. The two-point correlation function of the ionization field (middle) grows with time as well, peaking near the midpoint of reionization \cite{Lidz2008}. This rise and fall is reflected in the 21-cm power spectrum as well (right), here modeled in the ``saturated limit'' $\TS \gg \TCMB$, in which case only fluctuations in $\psi = \xHI (1 + \delta)$ need be considered. Larger values of $\zeta$ (thicker lines in right panel of Fig. \ref{fig:fzh04}) result in stronger fluctuations on large scales and a suppression in the power on small scales.

\begin{figure*}[]
\begin{center}
\includegraphics[width=0.98\textwidth]{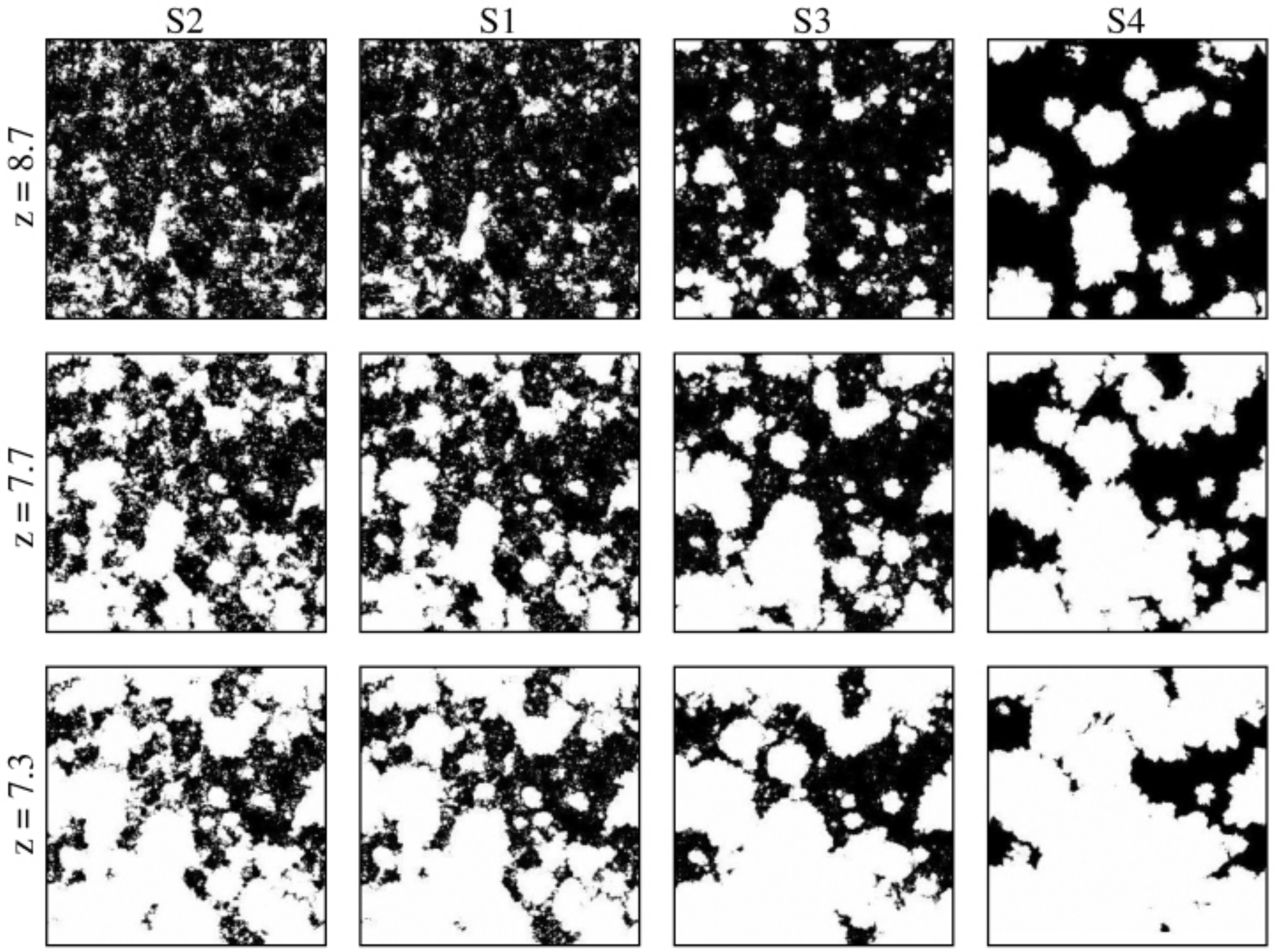}
\end{center}
\caption{{\bf Ionization field in $(94 \ \mathrm{cMpc})^3$ box for different ionizing efficiency assumptions \cite{McQuinn2007} as a function of redshift (top to bottom).} Neutral gas is black, while ionized regions are white.  Each column shows results from a different simulation, with S1-S3 using $\zeta=\mathrm{constant}$, $\zeta \propto m_h^{-2/3}$, and $\zeta \propto m_h^{2/3}$, respectively. S4 is the same as S1 except only halos with $m_h > 10^{10} \ M_{\odot}$ are included.}
\label{fig:mcquinn}
\end{figure*}

Figure \ref{fig:mcquinn} shows results from four different numerical simulations (RT post-processed on N-body simulation) \cite{McQuinn2007}, each differing in their treatment of $\zeta$. The key difference is how $\zeta$ depends on halo mass -- here, models span the range of $\zeta \propto m_h^{-2/3}$ (S2) to $\zeta \propto m_h^{2/3}$ (S3), including the case of $\zeta = \mathrm{constant}$ (S1). As the ionizing emissivity becomes more heavily weighted toward more massive, more rare halos (in S3 and S4), ionized structures grow larger and more spherical, while the smaller bubbles nearly vanish. This is a result of an increase in the typical bias of sources as $\zeta$ increases with $m_h$ -- because more massive halos are more clustered, ionizing photons from such halos combine to make larger ionized regions, whereas less clustered low-mass halos carve out smaller, more isolated ionized bubbles.

From Fig. \ref{fig:mcquinn} it is clear that the behavior of $\zeta$ not only sets the timeline for reionization but also its topology. However, $\zeta$ is degenerate with $\mmin$, since the ionizing emissivity can be enhanced both by increasing $\zeta$ directly or by increasing the number of star-forming halos by decrasing $\mmin$ (recall that the total number of ionizing photons emitted in a region is $N_{\gamma} = \zeta \fcoll$). Despite this degeneracy, power spectrum measurements expected to be able to place meaningful constraints on both parameters \cite{Greig2017}. The power spectrum on $k \sim 0.2 \ \mathrm{cMpc} h^{-1}$ scales reliably peaks near the midpoint of reionization \cite{Lidz2008}, meaning some (relatively) model-independent constraints are expected as well.

% f_X, alpha_X, Emin
\subsection{Dependence on the X-ray Efficiency and Spectrum} \label{sec:dep_xray}
The progression of cosmic reheating is analogous in some respects to reionization though driven by sources of much harder photons (see \S\ref{sec:hmxbs}-\ref{sec:ism_bremms}). As a result, we must consider the total energy emitted in X-rays (per unit collapsed mass or star formation rate) in addition to parameters that control the SED of X-ray sources. This is often achieved through a three parameter power-law model (see \S\ref{sec:xray_esc}), including a normalization parameter ($\zeta_X$), spectral cutoff ($E_{\min}$), and power-law slope of X-ray emission ($\alpha_X$). The combination of these parameters can capture a variety of physical models and mimic the shape of more sophisticated theoretical models (e.g., the multi-color disk spectrum; \cite{Mitsuda1984}).

\begin{figure*}[]
\begin{center}
\includegraphics[width=0.49\textwidth]{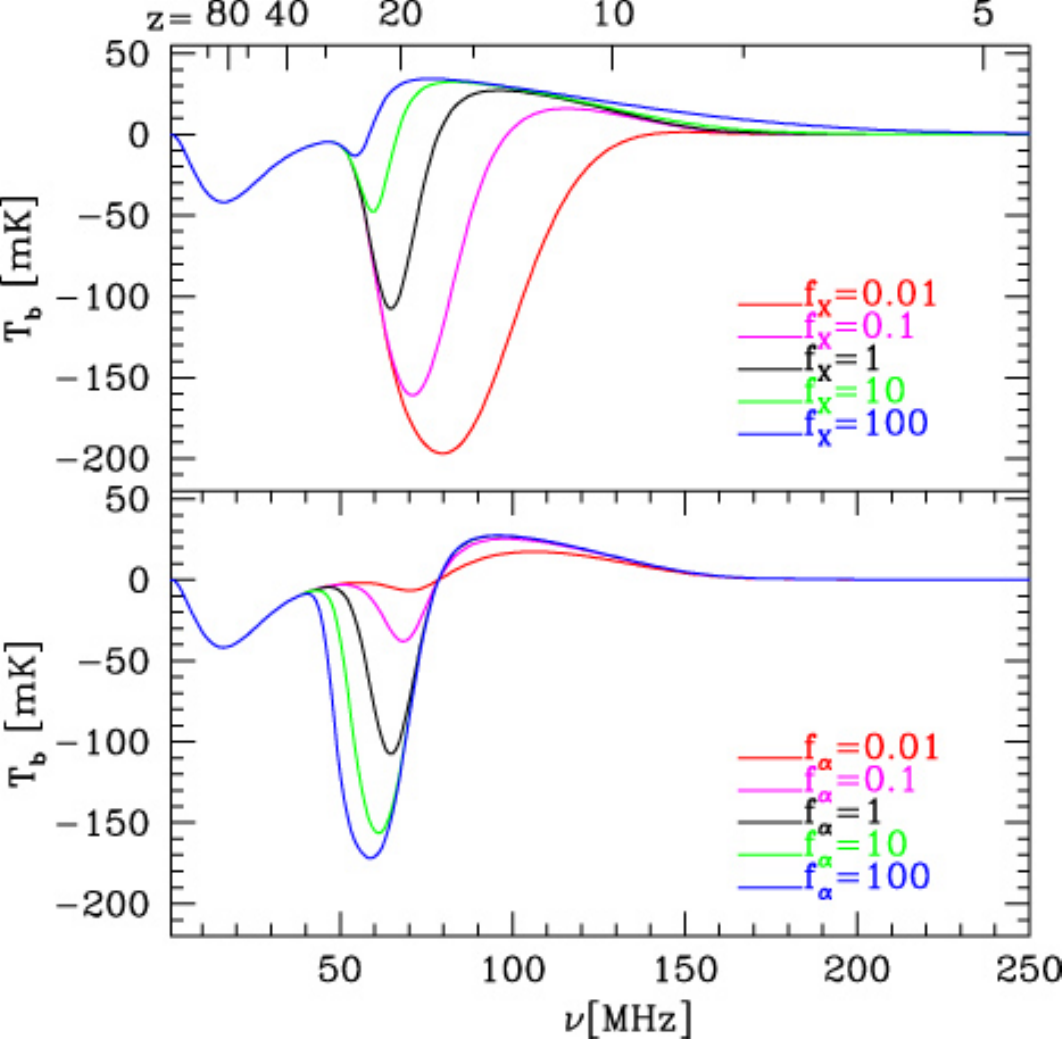}
\includegraphics[width=0.49\textwidth]{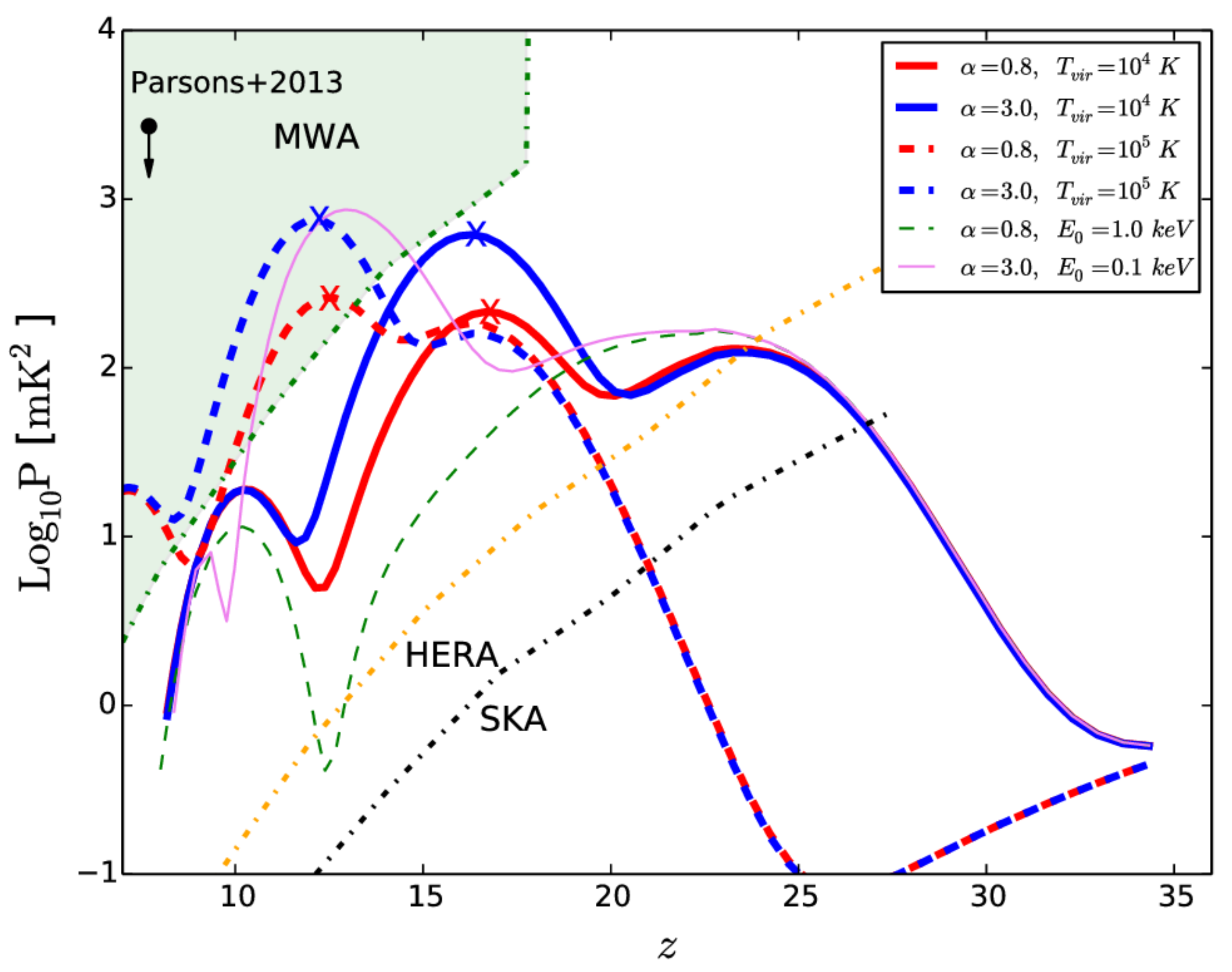}
\end{center}
\caption{{\bf Effects of $\Lya$ and X-ray efficiencies on the global 21-cm signal \cite{Pritchard2010,Pacucci2014}.} \textit{Left:} Predictions for the global 21-cm signal showing sensitivity to the normalization of the $L_X$-SFR relation, $f_X$ (top), and the production efficiency of $\Lya$ photons, $f_{\alpha}$ (bottom) \cite{Pritchard2010}. \textit{Right:} Predictions for evolution in the 21-cm power spectrum at $k=0.2 \ \mathrm{Mpc}^{-1}$ for models with different X-ray spectra \cite{Pacucci2014}. Blue curves indicate soft power-law spectra with indices of $\alpha=3$, while red curves are indicative of hard spectra sources with $\alpha=0.8$. Linestyles denote different minimum virial temperatures, $\Tmin$, and lower energy cutoffs for the X-ray background, $E_0$.}
\label{fig:fX}
\end{figure*}

Holding the SED fixed, variations in $\zeta_X$ affect the thermal history much like $\zeta$ affects the ionization history: increasing $\zeta_X$ causes efficient heating to occur earlier in the Universe's history, resulting in lower frequency (and shallower) absorption troughs in the global 21-cm signal, while the peak amplitude of the power spectrum also shifts to earlier times (at fixed wavenumber). Effects of $f_X$ on the global signal can be seen in the left panel of Figure \ref{fig:fX} (see also \cite{Mirabel2011,Mirocha2014,Fialkov2014b}).

Allowing the SED of X-ray sources can change the story dramatically because the mean free path of X-rays is a strong function of photon energy ($l_{\mathrm{mfp}} \propto \nu^{-3}$ due to the bound-free absorption cross section scaling $\sigma \propto \nu^{-3}$; \cite{Verner1996}). This strong energy dependence means that photons with rest energies $h\nu \gtrsim 2$ keV will not be absorbed within a Hubble length at $z \gtrsim 6$. Photons with $h\nu \lesssim 2$ keV will be absorbed on scales anywhere from $\sim 0.1$ Mpc to hundres of Mpc. As a result, the ``hardness'' of X-ray sources will determine the spatial structure of the kinetic temperature field -- soft photons will be absorbed on small scales and thus give rise to strong temperature fluctuations, while hard photons will travel great distances and heat the IGM more uniformly. The right panel of Figure \ref{fig:fX} shows precisely this effect -- for atomic cooling halos (solid lines), soft $\alpha_X = 3$ X-ray sources generate fluctuations $\sim 2-3$ times as strong as hard $\alpha_X=0.8$ sources \cite{Pacucci2014}, where in this case the spectral index is defined as $L_X \propto h\nu^{-\alpha_X}$.

The hardness of X-ray emission is also controlled by the ``cutoff energy,'' $E_{\min}$, below which X-ray emission does not escape efficiently from galaxies (see \S\ref{sec:xray_esc}). Alternatively, even if X-rays can escape efficiently, $E_{\min}$ could indicate an intrinsic turn-over in the X-ray spectra of galaxies, e.g., that expected from a multi-color disk spectrum \cite{Mitsuda1984}. Joint constraints on $\zeta_X$, $E_{\min}$, and $\alpha_X$ are thus required to help identify the sources of X-ray emission in the early Universe and the extent to which their spectra are attenuated by their host galaxies.

Finally, it is important to note that interpreting $\zeta_X$ is potentially more challenging than interpreting $\zeta$ given the additional parameters needed to describe the X-ray SED. One must be mindful of the fact that $\zeta_X$ quantifies the X-ray production efficiency \textit{in some bandpass}, generally 0.5-8 or 0.5-2 keV. As a result, changing $\alpha_X$ or $E_{\min}$ may be accompanied by a normalization shift so as to preserve the meaning of $\zeta_X$. This degeneracy can be mitigated to some extent by re-defining $\zeta_X$ in the $(E_{\min}-2$ keV band so as to isolate the photons most responsible for heating \cite{Greig2017}.

% N_alpha
\subsection{Dependence on the Ly-$\alpha$ Efficiency} \label{sec:dep_alpha}
The production efficiency of $\Lya$ photons affects when the 21-cm background first ``turns on'' due to Wouthuysen-Field coupling (see Ch. 1 and \S\ref{sec:lya}). Figure \ref{fig:fX} (bottom left panel) illustrates the effect increasing $f_{\alpha}$ has on the global 21-cm signal \cite{Pritchard2010}. For very large values, $f_{\alpha} = 100$ (blue), the dark ages come to an end at $z \sim 30$ ($\nu \sim 45$ MHz), triggering a much deeper absorption trough than the fiducial model (with $f_{\alpha} = f_X = 1$; black lines). The intuition here is simple: at fixed $f_X$, increasing $f_{\alpha}$ drives $\TS \rightarrow \TK$ at earlier times, meaning there has been less time for sources to heat the gas. 

Despite the very long mean-free paths of photons that source the $\Lya$ background, there are still fluctuations in the background intensity $J_{\alpha}$ \cite{Barkana2005,Ahn2009,Holzbauer2012}. As a result, there will be fluctuations in the spin temperature, as different regions transition from $\Ts \approx \TCMB$ to $\TS \approx \TK$ at different rates. The onset of $\Lya$ coupling is visible in the right panel of Figure \ref{fig:fX}, as the power (at k=0.2 $\mathrm{Mpc}^{-1}$) departs from its gradual descent at $z \sim 25$ ($\Tmin = 10^{5}$ K) and $z \sim 33$ ($\Tmin = 10^4$ K).

Because the $\Lya$ background is sourced by photons in a relatively narrow frequency interval, $10.2 \lesssim h \nu / \mathrm{eV} \lesssim 13.6$, the timing of Wouthuysen-Field coupling and the amplitude of fluctuations are relatively insensitive to the SED of sources. Similarly, because hydrogen gas is transparent to these photons (except at the $\Lyn$ resonances) these photons have an escape fraction $f_{\mathrm{esc,LW}} \gtrsim 0.5$, at least in the far field limit \cite{Schauer2015}, as their only impediment is $\Htwo$, which is quickly dissociated by stellar Lyman-continuum emission.

% Stellar properties
\subsection{Dependence on Stellar Metallicity}
As shown in the previous sections, it is common to allow $\zeta$, $\zeta_X$, and $\zeta_{\alpha}$ to vary independently as free parameters. However, if all features of the 21-cm background are driven by stars and their remnants, and the properties of such objects do not vary with time, then these efficiency factors will be highly correlated. For example, the number of Lyman continuum photons produced per unit star formation is inversely proportional to stellar metallicity, $Z$, as is the yield in the Lyman Werner band, so it may be more appropriate to use $Z$ as the free parameter instead of $\Nion$ and $\Nlw$. It is more difficult to connect the X-ray luminosity per baryon, $N_X$, to $Z$ as it depends on poorly understood details of the late stages of stellar evolution and compact binaries \cite{Belczynski2002}. However, observationally the $L_X$-SFR relation (see \S\ref{sec:hmxbs}) does appear to depend on gas-phase metallicity \cite{Brorby2016}, providing a simple empirical recipe for connecting $\zeta_X$ to $Z$ \cite{Mirocha2017}.

\begin{figure*}[]
\begin{center}
\includegraphics[width=0.98\textwidth]{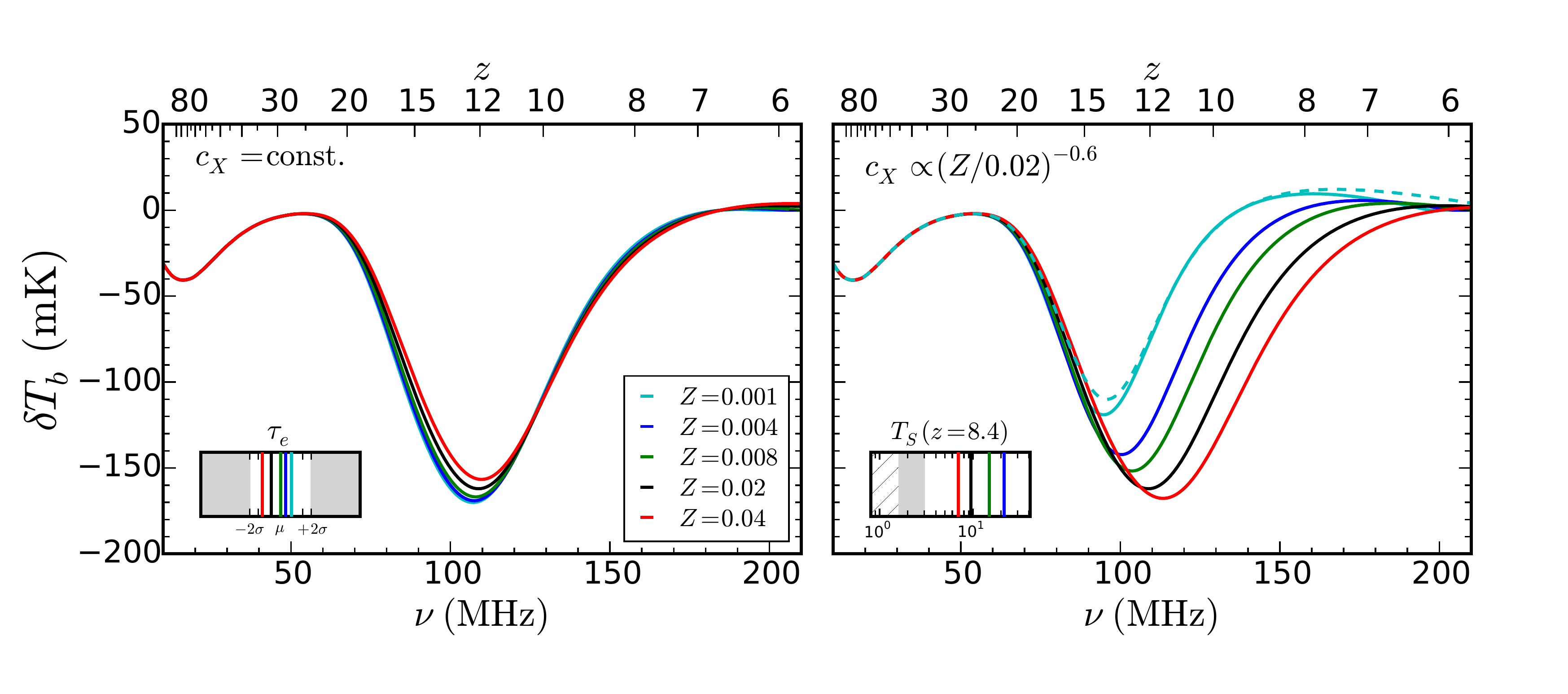}
\end{center}
\caption{{\bf Effects of stellar metallicity on the global 21-cm signal \cite{Mirocha2017}.} \textit{Left:} Metallicity effects assuming no link between stellar metallicity, $Z_{\ast}$, and X-ray luminosity. \textit{Right}: Metallicity effects assuming empirical relation between $f_X$ and $Z_{\ast}$ \cite{Brorby2016}. Insets show predictions for CMB optical depth, $\tau_e$ (left) and mean IGM spin temperature at $z=8.4$ (right).}
\label{fig:gs_metallicity}
\end{figure*}

Figure \ref{fig:gs_metallicity} shows these effects on the global 21-cm signal \cite{Mirocha2017}. In the left panel, no link between $L_X$/SFR and $Z$ is assumed, while in the right panel the empirical relation with $f_X \propto Z^{-0.6}$ is adopted. In each case, though particularly in the left panel, the effect of metallicity is very small. This is because these models force a match to high-$z$ UVLF measurements \cite{Bouwens2015,Finkelstein2015}, which means any change in $Z_{\ast}$ also affects the 1600 $\AA$ luminosity to which UVLF measurements are sensitive. As a result, changes in metallicity make galaxies more or less bright in the UV, but to preserve UVLFs, the efficiency of star formation must compensate (see \S\ref{sec:sfe}-\ref{sec:UV}). Once $f_X$ depends on $Z$ (right panel), the global 21-cm signal becomes more sensitive to changes in $Z$ because the change in X-ray luminosity can overcome the decline in SFE as $Z$ decreases \cite{Mirocha2017}.

In reality, the metallicity is a function of galaxy mass and time, so the simple constant $Z_{\ast}$ models above are of course simplistic. Note also that the models in Figure \ref{fig:gs_metallicity} only include atomic cooling halos. As a result, observed signals peaking at lower frequencies (like the EDGES 78 MHz signal \cite{Bowman2018}) likely require minihalos and/or non-standard source prescriptions \cite{Mirocha2018,Mirocha2019}.

% Mmin(z)
\subsection{Dependence on the Minimum Mass}
The minimum halo mass (or equivalent virial temperature) for star formation sets the total number of halos emitting UV and X-ray photons as a function of redshift and thus influences all points in 21-cm background, unlike the $\zeta$ factors, which largely impact a single feature. Fiducial models often adopt the mass corresponding to a virial temperature of $10^4$ K, since gas in halos of this mass will be able to cool atomically, i.e., there is not an obvious barrier to star formation in halos of this mass. Reducing $\mmin$, as is justified if star formation in minihalos is efficient, results in a larger halo population, while increasing $\mmin$ of course reduces the halo population. Moreover, for models in which low-mass halos are the dominant sources of emission, the typical star-forming halo is less biased than that drawn from a model in which high-mass halos dominate. As a result, changing $\mmin$ in principle affects both the timing of events in the 21-cm background as well as the amplitude of fluctuations. 

\begin{figure*}[]
\begin{center}
\includegraphics[width=0.98\textwidth]{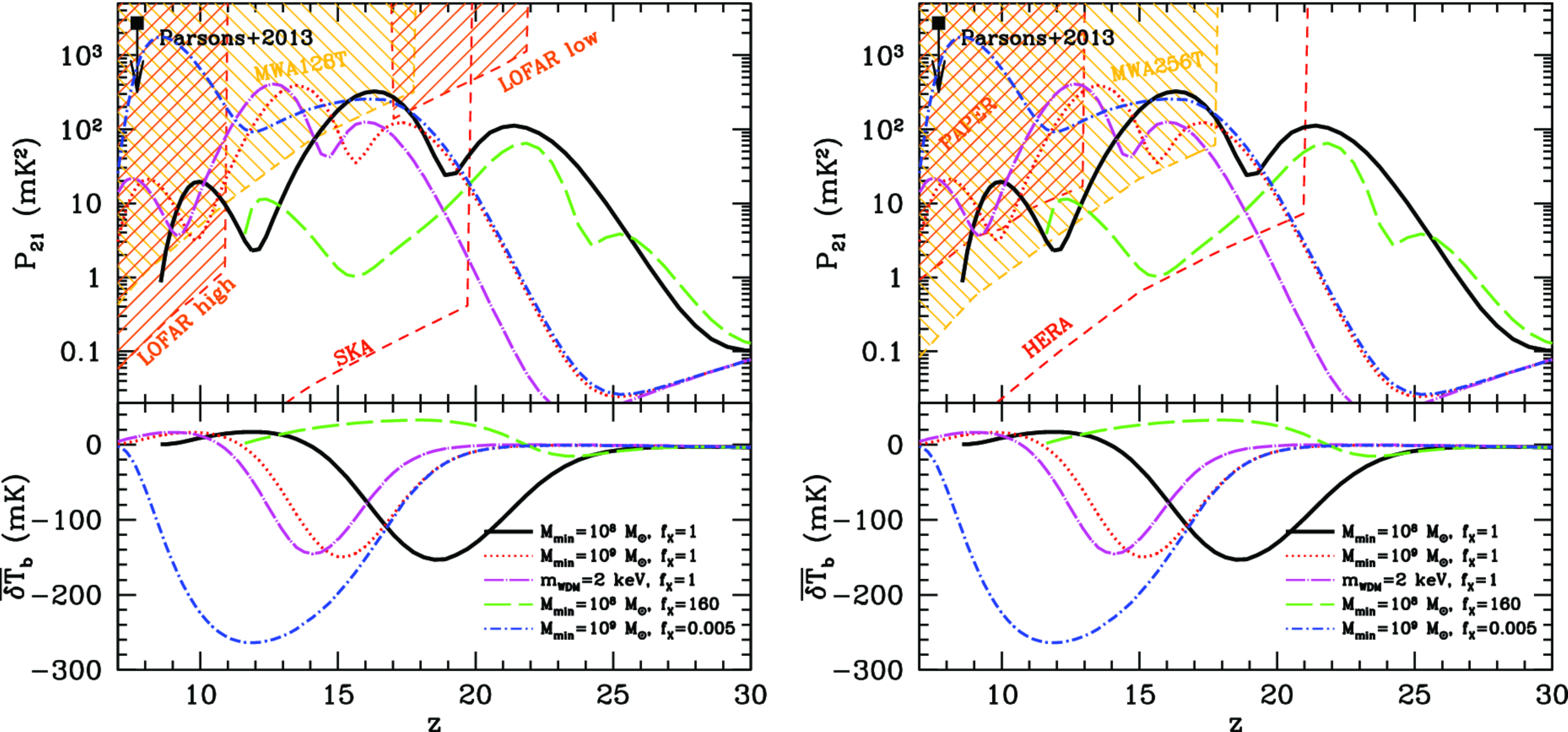}
\end{center}
\caption{{\bf Effects of the minimum mass on the global 21-cm signal and 21-cm power spectrum on $k=0.1 \ \mathrm{Mpc}^{-1}$ scales \cite{Mesinger2014}.} Solid black, red dotted, and magenta dash-dotted curves hold constant $f_X=1$ and vary the $\mmin$ by-hand (first two) and via a 2 keV warm dark matter particle (magenta). Left and right panels differ only in which sensitivity curves are included for comparison.}
\label{fig:mesinger2014_fig3}
\end{figure*}

As shown in Figure \ref{fig:mesinger2014_fig3} \cite{Mesinger2014}, $\mmin$ indeed affects all features of the 21-cm background, both in the global signal and fluctuations (see also, e.g., \cite{Fialkov2017,Mirocha2015}). With no other changes to the model the effects are largely systematic, i.e., the timing of features in the global signal and power spectrum are shifted without a dramatic change in their amplitude. Notice also that changing $\mmin$ can serve to mimic the effects of including warm dark matter (e.g., red dotted vs. magenta dash-dotted curve), which suppresses the formation of small structures that would otherwise (presumably) host galaxies.

Not depicted in Figure \ref{fig:mesinger2014_fig3} is the possibility that $\mmin$ evolves with time. Initially, only a mild redshift-dependence is expected just from linking $\mmin$ to a constant virial temperature of $\sim 500$ K \cite{Tegmark1997}, which is required for molecular cooling and thus star formation to initially begin (see \S\ref{sec:popIII}). However, the ability of minihalos to form stars also depends on their ability accrete and retain gas, which is influenced by the relative velocity between baryons and dark matter \cite{Tseliakhovich2010,Tseliakhovich2010,Fialkov2012}. Soon after the first sources form, $\mmin$ will react to the LW background \cite{Haiman1997,Machacek2001,Visbal2014}, and likely rise to the atomic cooling threshold, $\Tmin \sim 10^4$ K, at $z \gtrsim 10$ \cite{Trenti2009,Mebane2018}. During reionization, this threshold may grow even higher, as ionization inhibits halos from accreting fresh gas from which to form stars \cite{Gnedin2000,Noh2014,Yue2016}.

\begin{figure*}[]
\begin{center}
\includegraphics[width=0.98\textwidth]{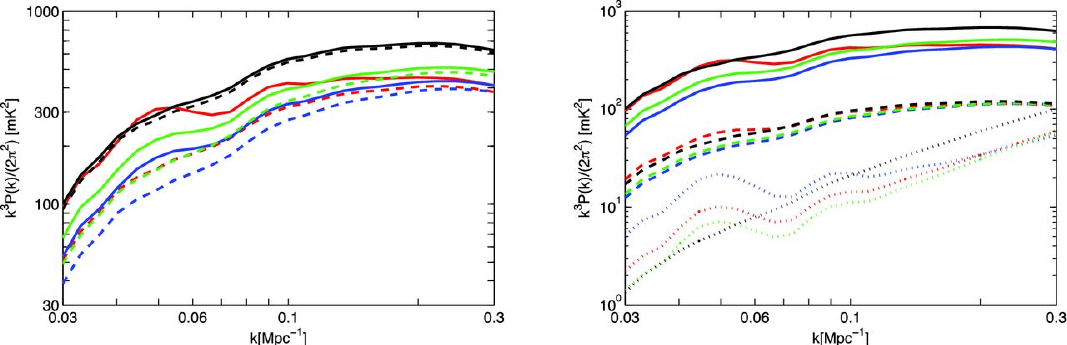}
\end{center}
\caption{{\bf Effects of LW feedback on the 21-cm power spectrum \cite{Fialkov2013}.} \textit{Left:} Power spectra with different feedback models, including no feedback (red), weak (blue), strong (green), and saturated (black). Dashed curves exclude the baryon-DM velocity offset effect \cite{Tseliakhovich2010}. \text{Right:} All models here include the baryon-velocity offset effect -- linestyles indicate power spectra at three different redshifts. Note that because changes to the strength of feedback shift the timing of events, models are compared at fixed increments relative to the ``heating redshift,'' $z_0$, which in these models occurs between $z\sim 15$ and $z\sim 18$ \cite{Fialkov2013}. Dashed lines are power spectra at $z=z_0$, while $z=z_0+3$ and $z=z_0+9$ are shown by solid and dotted lines, respectively.}
\label{fig:LWfeedback}
\end{figure*}

Figure \ref{fig:LWfeedback} shows 21-cm power spectra for various models of feedback in the first star-forming halos \cite{Fialkov2013}. Both the strength of feedback and type of feedback (LW and/or baryon-velocity streaming in this particular example) change the power spectrum by a factor of $\sim 2-3$ while fundamentally altering its shape. 

\begin{figure*}[]
\begin{center}
\includegraphics[width=0.98\textwidth]{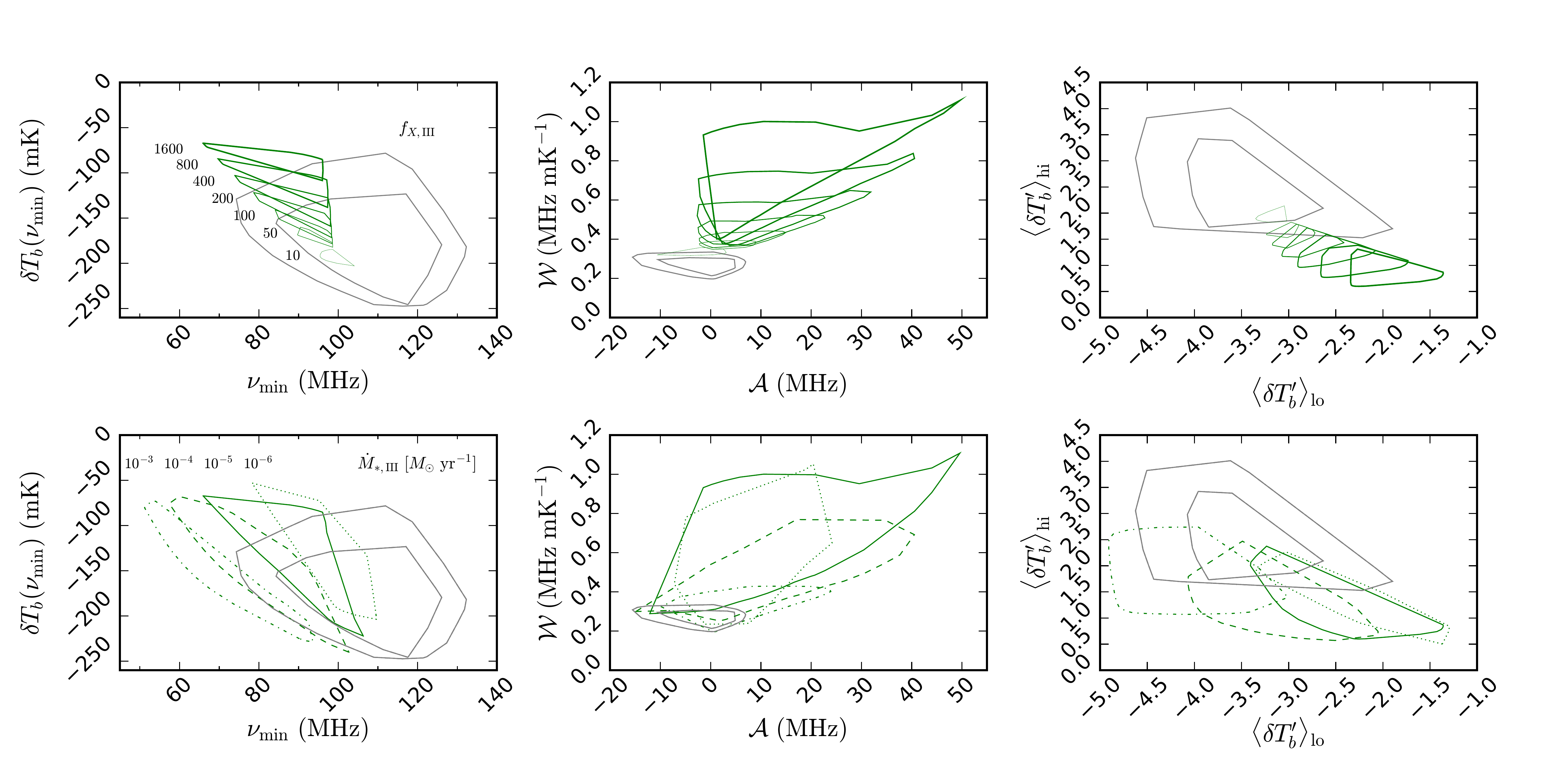}
\end{center}
\caption{{\bf Potential Pop~III signatures in the global 21-cm signal \cite{Mirocha2018}.} Comparison between Pop~II-only and Pop~III models performed in three diagnostic spaces, including the absorption trough position (left), the prominence of its wings, $\mathcal{W}$, and its asymmetry, $\mathcal{A}$ (middle), and the mean slopes at frequencies above and below the extremum (right). Black contours enclose sets of PopII models generated by Monte Carlo sampling a viable range of parameter space constrained by current observations, while the green polygons are slices through a 3-D Pop~III model grid, first assuming assuming a fixed Pop~III SFE and varying the X-ray production efficiency (top row), and then for different SFE models having ‘marginalized’ over all $f_{X,\textsc{iii}}$ (bottom row). Measurements falling in regions of overlap between the green and black contours would have no clear evidence of PopIII, while measurements falling only within the green contours would be suggestive of PopIII.}
\label{fig:popIII_gs}
\end{figure*}

The signatures of minihalos and feedback in the global 21-cm signal are likely more subtle. Figure \ref{fig:popIII_gs} shows predictions for the amplitude and shape of the global 21-cm absorption signal with (green) and without (gray) a model for Pop~III star formation and LW feedback \cite{Mirocha2018}. While the effects of Pop~III sources on the position of the absorption trough alone are difficult to distinguish from uncertainties in Pop~II source models (quantified by gray contours), as shown in the left column, Pop~III signals do affect the symmetry of the trough and the derivative of the signal (middle and right, respectively). As a result, any inferred skew in the global signal (to high frequencies) may be an indicator of efficient Pop~III star formation in the early Universe.

%\begin{center}
%\begin{table}
%\begin{tabular}{||c | c | c||}
%\hline
%name & description & typical values \\ 
%\hline\hline
%$\zeta_i$ & Ionizing photon production efficiency & 40 ish  \\ 
%\hline
%$\zeta_{\alpha}$ & $\Lya$ photon production efficiency & 40 ish  \\ 
%\hline
%$\zeta_X$ & X-ray photon production efficiency & xxx \\
%\hline
%$\Tmin$ & Minimum virial temperature of star-forming halos & $10^4$ K \\
%\hline
%\end{tabular}
%\caption{Parameters in simple 21-cm models.}
%\end{table}
%\end{center}
%

% Sinks
%\subsection{Sources v. Sinks}
%{\color{red} include this?}

%%
% Summary?
%%
\section{Summary}
In this chapter we have introduced the fundamentals of ionization and heating in the high-$z$ IGM (\S\ref{sec:igm}), the sources most likely to provide the UV and X-ray photons that drive reionization and reheating (\S\ref{sec:sources}), and the signatures of these events in the global 21-cm signal and power spectrum (\S\ref{sec:predictions}). However, this is by no means an all-inclusive account of work in the field. There are a variety of modeling codes \cite{Furlanetto2004,Barkana2005,Pritchard2007,Thomas2009,Mesinger2011,Fialkov2014a,Mirocha2014,Santos2010,Raste2018}, each designed with different trade-offs in mind and often with their own methods for parameterizing the effects of astrophysical sources. To date, no systematic effort has been undertaken to compare the results of these codes or to (attempt to) converge on a ``concordance'' parameterization of the cosmic dawn. As a result, we encourage readers to be aware of the assumptions underlying different models, how parameters are defined, and extent to which these choices impact the inferences drawn from current and future experiments.

\bibliographystyle{plain}
\bibliography{References}

%\include{Furlanetto/chapter}
%\include{chapter}
%\include{Pritchard/chapter}
%\include{Greig/chapter}
%\include{Bernardi/chapter}
%\include{Chapman_Jelic/chapter}
%\include{Greenhill_Subrahmanyan/chapter}
%\include{Trott_Pober/chapter}
%\include{Koopmans_Bernardi/chapter}

%\part{Theory}
%\appendix

\end{document}